\documentclass[a4paper,11pt,reqno]{article}
\usepackage{a4wide}
\usepackage{amsmath,amsfonts,amssymb,txfonts,pxfonts}
\usepackage[english]{babel}
\usepackage{soul}
\usepackage{nicefrac}
\usepackage[mathscr]{euscript}
\usepackage{setspace}
\usepackage{datetime}
\usepackage[nosort]{cite}
\usepackage{mathrsfs}
\usepackage[T1]{fontenc}

\usepackage{txfonts}
\usepackage{kpfonts}
\usepackage{upgreek}

\usepackage{comment}

\usepackage{mathpazo}
\usepackage{graphicx}
\usepackage[breaklinks=true]{hyperref}

\numberwithin{equation}{section}

\hypersetup{
			colorlinks=true,
			urlcolor=blue,
			citecolor=magenta,
			linkcolor=blue,
     }

\let\oldsqrt\sqrt
\def\sqrt{\mathpalette\DHLhksqrt}
\def\DHLhksqrt#1#2{%
\setbox0=\hbox{$#1\oldsqrt{#2\,}$}\dimen0=\ht0
\advance\dimen0-0.2\ht0
\setbox2=\hbox{\vrule height\ht0 depth -\dimen0}%
{\box0\lower0.4pt\box2}}

\def\crbig{\\\noalign{\vspace{1.1mm}}}
\newcommand{\RNum}[1]{\uppercase\expandafter{\romannumeral #1\relax}}

\author{
  \begin{minipage}{.97\linewidth}
    \vspace{1cm}
       \begin{center}
      \begin{small}
               \textbf{Luca Ciambelli},$^1$  
               \textbf{Charles Marteau},$^1$ 
             \textbf{Anastasios C. Petkou},$^{2,3}$ 
                     \\
     \textbf{P. Marios Petropoulos}$^1$ and 
      \textbf{Konstantinos Siampos}$^{3,4}$
              \end{small}
    \end{center}
    \vspace{0.5cm}
    \hspace{2.4cm}\begin{minipage}{.7\linewidth}
\begin{center}     {\it \begin{footnotesize}
\hbox{\kern-1.8cm\vbox{\vskip0cm
 \begin{itemize}
               \item[$^1$]CPHT -- Centre de Physique Th\'eorique\\ 
        Ecole Polytechnique, CNRS UMR 7644\\
        Universit\'e Paris--Saclay\\
        91128 Palaiseau Cedex, France
\vskip0.3cm
      \end{itemize}}
\kern-3cm\vbox{
\begin{itemize}
 \item[$^2$]Department of Physics\\ 
  Institute of Theoretical Physics\\
  Aristotle University of Thessaloniki\\ 
  54124 Thessaloniki, Greece
      \end{itemize}
      \vskip0.05cm
}}
     \end{footnotesize}}
\end{center}
    \end{minipage}
    \vspace{0.5cm}\begin{minipage}{.7\linewidth}
\begin{center}     
{\it \begin{footnotesize}
\hbox{\kern0.6cm\vbox{\vskip0cm
 \begin{itemize}
              \item[$^3$] Theoretical Physics Department\\
CERN\\ 
1211 Geneva 23, Switzerland
\vskip0.45cm
      \end{itemize}}
\kern-3cm\vbox{
\begin{itemize}
 \item[$^4$] Albert Einstein Center for Fundamental Physics\\
Institute for Theoretical Physics\\ 
University of Bern\\
Sidlerstrasse 5, 3012 Bern, Switzerland
      \end{itemize}\vskip0.05cm
}
}
     \end{footnotesize}}
\end{center}
     \end{minipage}
  \end{minipage}
}

\title{\vspace{1.5cm}
 \boldmath \begin{Large}
    \textbf{Covariant Galilean versus Carrollian hydrodynamics from relativistic fluids}
  \end{Large} \unboldmath
}

\date{}

\begin{document}

\begin{titlepage}
\maketitle
\thispagestyle{empty}

 \vspace{-14.cm}
  \begin{flushright}
  CPHT-RR048.082017\\
CERN-TH-2017-228
  \end{flushright}
 \vspace{12.cm}

\begin{center}
\textsc{Abstract}\\  
\vspace{0.8 cm}	
\begin{minipage}{1.0\linewidth}

We provide the set of equations for non-relativistic fluid dynamics on arbitrary, possibly time-dependent  spaces, in general coordinates. These equations are fully covariant under either local Galilean or local Carrollian transformations, and are obtained from standard relativistic hydrodynamics in the limit of infinite or vanishing velocity of light. All dissipative phenomena such as friction and heat conduction  are included in our description. Part of our work consists in designing the appropriate coordinate frames for relativistic spacetimes, invariant under Galilean or Carrollian diffeomorphisms. The guide for the former is the dynamics of relativistic point particles, and leads to the Zermelo frame. For the latter, the relevant objects are relativistic instantonic space-filling branes in Randers--Papapetrou backgrounds. We apply our results for obtaining the general first-derivative-order Galilean fluid equations, in particular for incompressible fluids (Navier--Stokes equations) and further 
illustrate our findings with two applications: Galilean fluids in rotating frames or inflating surfaces and Carrollian conformal  fluids on two-dimensional time-dependent geometries. The first is useful in atmospheric physics, while the dynamics emerging in the second is governed by the Robinson--Trautman equation, describing a Calabi flow on the surface, and known to appear when solving Einstein's equations for algebraically special { Ricci-flat or Einstein} spacetimes.

\end{minipage}
\end{center}

%\vspace{2cm} 

\end{titlepage}

\onehalfspace

%\noindent\rule{\textwidth}{1.2pt}
%\vspace{-1cm}
\begingroup
\hypersetup{linkcolor=black}
\tableofcontents
\endgroup
\noindent\rule{\textwidth}{0.6pt}

\section{Introduction}

Ordinary non-relativistic fluid dynamics is described in terms of a basic set of equations: continuity, energy conservation and momentum conservation (Euler equation). In most textbooks (as \emph{\emph{e.g.}} \cite{Landau}) the fluid is observed from either inertial, or stationary rotating frames, using Cartesian or spherical/cylindrical coordinates. Although these set-ups are satisfactory for most  practical purposes, they do not exhaust all possible situations because the equations at hand are not covariant under Galilean diffeomorphisms \emph{i.e.} general coordinate transformations such as $t'=t'(t)$ and $\textbf{x}^{\prime}=\textbf{x}^{\prime}(t, \textbf{x})$. Most importantly, the geometry hosting the fluid is assumed to be three- or two-dimensional Euclidean space. This is a severe limitation, as we may want to study the fluid moving  on a surface, which is neither flat nor static, and equipped with an arbitrary coordinate system. 

Progress has been made over the last decades, sustained by the needs of the space programs or meteorology \cite{Vinokur, Avis, Aris, Carlson, Luo, Charron}. The most recent work \cite{Charron} beautifully highlights  the various contributions, and provides a covariant frame-independent formulation. Still, these authors do not address the issue of trading Euclidean space for an arbitrary curved and time-dependent geometry, { and subsequent analyses have focused to the case of static surfaces (see \emph{e.g.} \cite{Debus})}. Part of our work consists in filling this gap, and presenting the  most general equations describing a non-relativistic viscous fluid moving on a space endowed with a spatial, time-dependent metric, and observed from an arbitrary frame. Each geometric object involved in this description has a well-defined transformation rule under Galilean diffeomorphisms, making the set of equations covariant. 

In order to achieve the above program, we carefully analyze the infinite-light-velocity limit inside the relativistic fluid equations. Although standard (see \textsection 125 of \cite{Landau} for the original presentation and \cite{RZ} for a modern approach), this method has been only partially developed  outside the realm of Minkowski spacetime (as \emph{e.g.} in \cite{Karch}). Hence, it has mostly led to non-relativistic fluids on plain Euclidean space in inertial frames. Choosing the form of a general spacetime
metric such that it allows for a non-relativistic limit, enables us to reach our goal. 

Considering the infinite-light-velocity limit in a relativistic framework suggests to study in parallel the alternative zero-light-velocity limit. This is actually ultra-relativistic, but we will keep on calling it non-relativistic as it decouples time and contracts the Poincar\'e group down to the Carroll group, as originally described in \cite{Levy}. 

Carrollian physics has attracted some attention over the recent years
\cite{Duval:2014uoa, Duval:2014uva}. Although kinematically restricted -- due to the vanishing velocity of light, the light-cone collapses to a line and no motion is allowed -- the freedom of choosing a frame is as big as for Galilean physics though. In particular, 
the single particle has degenerate motion \cite{Bergshoeff:2014jla}, but extended 
instantonic\footnote{In ordinary relativistic spacetime, we would call these objects tachyonic as they extend in space \emph{i.e.} outside the local light-cone. Since the latter is everywhere degenerate in Carrollian spacetimes, instantonic is more illustrative. \label{tachin}}
 objects do still exist and have non-trivial dynamics, making this framework rich and
interesting. Following the pattern described above, we study the corresponding general set of equations for viscous fluids. The form of the spacetime metric appropriate for the limit at hand is of  Randers--Papapetrou, slightly different from the one used in the former case, which is the Zermelo form.\footnote{See \cite{Gibbons:2008zi} for an interesting discussion on Zermelo
 vs.  Randers--Papapetrou forms.} The obtained equations are covariant under Carrollian coordinate transformations, $t'=t'(t, \textbf{x})$ and $\textbf{x}^{\prime}=\textbf{x}^{\prime}(\textbf{x})$. In order to avoid any confusion, we will refer to the standard non-relativistic fluids as \emph{Galilean}, whereas the latter will be called  \emph{Carrollian}. 

Our motivation for the present work is twofold. On the one hand, as already mentioned, stands the need for a fully covariant formulation of Galilean fluid dynamics, on general spaces and from arbitrary frames, which might have useful physical applications. On the other hand, viscous Carrollian fluids were never studied and turn out to emerge in the context of asymptotically flat holography \cite{CMPPS2}, in replacement of the relativistic fluids present in the usual fluid/gravity holographic correspondence of asymptotically anti-de Sitter spacetimes  \cite{Bhattacharyya:2007, Haack:2008cp, Bhattacharyya:2008jc, Hubeny:2011hd}. Performing this analysis in parallel is useful as both Galilean and Carrollian groups, and Zermelo and Randers--Papapetrou frames turn out to have intimate duality relationships. 

We will start our exposition by designing the appropriate forms for relativistic spacetimes, hosting naturally the action of -- \emph{i.e.} being stable under -- the two diffeomorphism groups that we want to survive in the infinite-$c$ or zero-$c$ limits,  Secs. \ref{GALdyn},  \ref{CARdyn}. Local Galilean and  
Carrollian transformations are elegantly implemented in ordinary particle or instantonic space-filling brane dynamics, respectively. They are subsequently uplifted 
into Zermelo and Randers--Papapetrou metrics for the spacetime. The next step consists in studying ordinary viscous relativistic fluids on these environments and consider the infinite-$c$ or zero-$c$ limits in their equations. This is performed in Secs.  \ref{FLUIDS} and \ref{FLUID2}, following a concise overview on relativistic fluids, Sec.  \ref{recap}. We find generalized continuity, energy-conservation and Euler equations for the usual Galilean fluids, as well as a set of two scalar (one for the energy) and two vector equations for the Carrollian ones. We analyze the covariance properties of the equations in both cases, and show that these transform as expected. Some examples are collected in Sec. \ref{EXA}: the Galilean fluid from a rotating frame or on an inflating surface, and the dynamics of a two-dimensional Carrollian viscous fluid. Further technical details, are provided in the appendix, where we introduce a new time connection for the Galilean geometry, and both temporal and spatial  
connections for the Carrollian and conformal-Carrollian geometry, together with their associated curvature tensors, allowing for a more elegant presentation of the corresponding covariant equations.

\section{Galilean and Carrollian Poincar\'e uplifts}

We present here the relativistic uplifts of Newton--Cartan and Carrollian non-relativistic structures. In these Lorentzian-signature spacetimes, respectively of the Zermelo and Randers--Papapetrou form, the Galilean and Carrollian diffeomorphisms are naturally realized, and the dynamics of free objects smoothly matches the ordinary Galilean and Carrollian dynamics, when the velocity of light becomes infinite or vanishes, respectively.

\subsection{From Galileo Galilei \dots}\label{GALdyn}

Consider a free particle on an arbitrary $d$-dimensional space $\mathscr{S}$, endowed with a positive-definite metric  
\begin{equation}
\label{dmet}
\text{d}\ell^2=a_{ij} \text{d}x^i \text{d}x^j,\quad i,j\ldots \in \{1,\ldots,d\},
\end{equation}
and observed from a frame with respect to which
the locally inertial frame has velocity $\mathbf{w}=w^i \partial_i$. 
Its classical (as opposed to relativistic) dynamics
is captured by the following Lagrangian:
\begin{equation}
\label{gallag}
\mathcal{L}(\mathbf{v},\mathbf{x},t)=\frac{1}{2\Omega^2}a_{ij} \left(v^i-w^i \right)\left(v^j-w^j\right)
\end{equation}
with action
\begin{equation}
\label{galactgal}
S[\mathbf{x}]=\int_{\mathscr{C}}\text{d}t\, \Omega \mathcal{L}(\mathbf{v},\mathbf{x},t).
\end{equation}

In this expression:
\begin{itemize}
\item $a_{ij}$ and $w^i$ are general functions of $(t,\mathbf{x})$;\footnote{Here $\mathbf{x}$ stands for $\{x^1,\ldots,x^d\}$.}
\item $v^i=\frac{ \text{d}x^i}{ \text{d}t}$ are the usual  components  of the velocity  $\mathbf{v}=v^i \partial_i$;
\item $\mathcal{L}(\mathbf{v},\mathbf{x},t)$ appears as a Lagrangian density, with Lagrangian\footnote{Euler--Lagrange equations are $\frac{\text{d}}{\text{d}t}\left(\frac{\partial L}{\partial v^i}\right)=\frac{\partial L}{\partial x^i}$.} 
 $L(\mathbf{v},\mathbf{x},t)=\Omega\mathcal{L}(\mathbf{v},\mathbf{x},t)$.
\end{itemize}
Furthermore 
\begin{itemize}
\item  the Lagrange generalized momenta are (indices are lowered and raised with $a_{ij}$ and its inverse)
\begin{equation}
\label{galmom}
p_i=\frac{\partial L}{\partial v^i}=\frac{1}{\Omega}(v_i-w_i),
\end{equation}
\item $H(\mathbf{p},\mathbf{x},t)=p_i v^i -L(\mathbf{v},\mathbf{x},t)$ is the Hamiltonian with Hamiltonian density $\mathcal{H}=\frac{1}{\Omega}H$:
\begin{equation}
\label{galenmom}
\mathcal{H}
=\frac{1}{2} \left(\mathbf{p}^2
+\frac{\mathbf{p}\cdot\mathbf{w}}{\Omega}
\right).
\end{equation}
\end{itemize}
The existence of an 
absolute Newtonian time requires $\Omega$ be a function of $t$ only, the absolute time being thus $\int \text{d}t\, \Omega(t) $. One should stress that keeping general $\Omega(t,\mathbf{x})$ does not spoil the consistency of the system \eqref{gallag}, \eqref{galactgal}, but invalidates the interpretation of \eqref{dmet} as the spatial metric. Even though in practical situations we can set $\Omega=1$, its r\^ole is important when dealing with general Galilean diffeomorphisms (see  \eqref{galdifa}--\eqref{galdifom}), in the framework underlying the above dynamical system: the
 Newton--Cartan structures \cite{NC}.\footnote{Some modern references on Newton--Cartan structure are \emph{e.g.} \cite{Duv, Bekaert:2014bwa, Bekaert:2015xua, Festu}.} 

We can compute the energy density expressing the Hamiltonian \eqref{galenmom}
 in terms of the velocity: 
\begin{equation}
\label{galen}
\mathcal{H}=\frac{1}{2\Omega^2}a_{ij} \left(v^i+w^i \right)\left(v^j-w^j\right)
=\frac{1}{2\Omega^2} \left(\mathbf{v}^2
-\mathbf{w}^2
\right).
\end{equation}
As usual $\nicefrac{-\mathbf{w}^2}{2\Omega^2} $ plays the r\^ole of the potential for inertial forces.
Using the energy theorem ($\nicefrac{\text{d}H}{\text{d}t}=\nicefrac{-\partial L}{\partial t}$) one finds
\begin{equation}
\label{galenth}
\frac{\text{d}\mathcal{H}}{\text{d}t}=
-\frac{1}{2\Omega^2} \left(v^i-w^i \right)\left(v^j-w^j\right)\partial_t a_{ij}
+\frac{v_i-w_i }{\Omega}\partial_t\frac{w^i}{\Omega}.
\end{equation}

The most canonical example of \eqref{gallag} is that of a massive particle moving in Euclidean space $E_3$ with Cartesian coordinates, and observed from a non-inertial frame: 
\begin{equation}
\label{E3-rot}
a_{ij}=\delta_{ij},\quad \Omega = 1,\quad 
\mathbf{w}(t,  \mathbf{x})= \mathbf{x}\times\pmb{\omega}(t)-\mathbf{V}(t).
\end{equation}
Here $\mathbf{V}(t)$ is the dragging velocity of the non-inertial frame, $\pmb{\omega}(t)$ the angular velocity of its rotating axes, and $\mathbf{v}-\mathbf{w}= \mathbf{v}+\mathbf{V}+\pmb{\omega}\times \mathbf{x}$ is the velocity as measured in the original inertial frame (Roberval's theorem). 

The action \eqref{galactgal} is invariant
under general \emph{Galilean diffeomorphisms i.e.}  transformations
\begin{equation}
\label{galdifs} 
t'=t'(t)\quad \text{and} \quad \textbf{x}^{\prime}=\textbf{x}^{\prime}(t, \textbf{x}),
\end{equation}
for which we define the following Jacobian functions:
\begin{equation}
 \label{galj}
J(t)=\frac{\partial t'}{\partial t},\quad j^i(t,\mathbf{x}) = \frac{\partial x^{i\prime}}{\partial t},\quad 
J^i_j(t,\mathbf{x}) = \frac{\partial x^{i\prime}}{\partial x^{j}}.
\end{equation}
The metric components transform as a tensor of $\mathscr{S}$:
\begin{equation}
\label{galdifa}
a^{\prime}_{ij} =a_{kl} J^{-1k}_{\hphantom{-1}i} J^{-1l}_{\hphantom{-1}j} ,
\end{equation}
the particle and frame velocities as gauge connections: 
\begin{eqnarray}
\label{galdifv}
v^{\prime k}&=&\frac{1}{J}\left(J^k_i v^i+j^k\right),
\\
\label{galdifw}
w^{\prime k}&=&\frac{1}{J}\left(J^k_i w^i+j^k\right),
\end{eqnarray}
and the generalized momenta \eqref{galmom} as one-form components:
\begin{equation}
\label{galdifp}
p^{\prime}_{i} =p_k J^{-1k}_{\hphantom{-1}i};
\end{equation}
$\Omega$ is just rescaled:
\begin{equation}
\label{galdifom}
\Omega^{\prime }=\frac{\Omega}{J}.
\end{equation}
Since $J=J(t)$ and $\Omega=\Omega(t)$, Galilean transformations  lead to $\Omega^\prime=\Omega^\prime(t^\prime)$, leaving invariant the absolute Newtonian time $\int \text{d}t\, \Omega(t) =\int \text{d}t^\prime\, \Omega^\prime(t^\prime) $.
Observe also that $\frac{\mathbf{v}-\mathbf{w}}{\Omega}$ is a genuine vector of $\mathscr{S}$, which ensures the form-invariance of $\mathcal{L}$ and thus the covariance of the equations of motion.

There is a particular Newton--Cartan structure, which is invariant under the Galilean group: $\mathscr{S}$ is the Euclidean space $E_d$ with Cartesian coordinates ($a_{ij}=\delta_{ij}$) and $\Omega = 1$, and the connection $\mathbf{w}$ is constant \emph{i.e.} independent of $(t, \textbf{x})$. 
This system describes the non-relativistic motion of a free particle in Euclidean space, observed from an inertial frame. The Galilean group acts as
\begin{equation}
\label{gal}
\begin{cases}
t'=t+t_0,\\
x^{\prime k}=R^k_i x^i+V^k t +x^k_0
\end{cases}
\end{equation}
with all parameters being $(t, \textbf{x})$-independent, and $R^k_i $ the entries of an orthogonal matrix.  The action of these transformations leave the Lagrangian and the equations of motion at hand \emph{invariant}. In more general Newton--Cartan structures, the Galilean group acts in the tangent space equipped with a local orthonormal frame and it is no more a global symmetry.

The Galilean group is an  infinite-$c$ contraction of the Poincar\'e group. The latter acts locally in general $d+1$-dimensional pseudo-Riemannian manifolds $\mathscr{M}$. In order to recover the above Newton--Cartan structure and its class of diffeomorphisms \eqref{galdifs} in the infinite-$c$ limit, there is a natural choice for the form of the metric on $\mathscr{M}$:
\begin{equation}
\label{galzerm}
\text{d}s^2 =-\Omega^2 c^2 \text{d}t^2+a_{ij} \left(\text{d}x^i -w^i  \text{d}t\right)\left( \text{d}x^j-w^j \text{d}t\right).
\end{equation}
The form \eqref{galzerm} is required for the functions $\Omega$, $a_{ij}$ and $w^i$ to transform as in  \eqref{galdifa}, \eqref{galdifw} and \eqref{galdifom} under a Galilean diffeomorphism \eqref{galdifs}. Actually, every metric is compatible with the gauge \eqref{galzerm}, provided  $a_{ij}$, $w^i$ and $\Omega$,  are free to depend on $x=(ct,\mathbf{x})=\{x^\mu, \mu=0,1,\ldots,d\}$. The existence of a Galilean limit requires, however, $\Omega$ to depend on $t$ only. Indeed, the proper time element for a physical observer  is $\text{d}\tau =\sqrt{\nicefrac{-\text{d}s^2}{c^2}}$. When $c$ becomes infinite, ${\lim\limits_{c\to \infty }\text{d}\tau=\Omega\, \text{d}t}$ must coincide with the absolute Newtonian time, and this requires the absence of $\mathbf{x}$-dependence in $\Omega$, as expected from our previous discussion on the dynamics  of \eqref{galactgal}.

The spacetime Jacobian matrix 
associated with \eqref{galdifs}, reads (using \eqref{galj}):
 \begin{equation}
 \label{galjst} 
J^\mu_\nu(x) = \frac{\partial x^{\mu\prime}}{\partial x^{\nu}}\to
\begin{pmatrix}   J(t) &0\\  
J^i(x) & J^i_j(x)
 \end{pmatrix} \quad \text{with}\quad J^i=\frac{j^i}{c} .
\end{equation}
The metric form \eqref{galzerm} is refered to as \emph{Zermelo} (see \cite{Gibbons:2008zi}). A relativistic particle moving in \eqref{galzerm} is described by the components of its velocity $\text{u}$, normalized as $\| \text{u} \|^2=-c^2$:
\begin{equation}
\label{vel}
u^\mu=\frac{ \text{d}x^\mu}{\text{d}\tau}\Rightarrow
u^0=\gamma c,\ u^i = \gamma v^i,
\end{equation}
where the Lorentz factor $\gamma$ is defined as usual (although here, it depends also on the spacetime coordinates):\footnote{Expressions as $\mathbf{v}^2$ stand for $a_{ij}v^i v^j$, not to be confused with $\| \text{u} \|^2=g_{\mu\nu}u^\mu u^\nu$.}
\begin{equation}
\label{gamzerm}
\gamma(t,\mathbf{x},\mathbf{v})=\frac{\text{d}t}{\text{d}\tau}=
\frac{1}{\Omega\sqrt{1-\left(\frac{\mathbf{v}-\mathbf{w}}{c\Omega}\right)^2}}.
\end{equation}
Under a Galilean diffeomorphism \eqref{galjst}, the transformation of the components of $\text{u}$,
\begin{equation}
u^{\prime 0}=J u^0,\quad u^{\prime i} = J^i_k u^k+J^i u^0,\quad
u^{\prime}_{0}=\frac{1}{J}\left( u_0-u_j J^{-1j}_{\hphantom{-1}k}
J^k \right),\quad u^{\prime}_{i} =u_k J^{-1k}_{\hphantom{-1}i} ,
\end{equation}
 induces a transformation on $v^i$, which matches precisely \eqref{galdifv}.

The dynamics of the relativistic free particle is described using \emph{e.g.} the length of the world-line $\mathscr{C}$ as an action:
\begin{equation}
\label{galact}
S[ x]=\int_{\mathscr{C}}\text{d}\tau =\int_{\mathscr{C}}\sqrt{-\frac{\text{d}s^2}{c^2}}.
\end{equation}
This is easily computed in the Zermelo environment \eqref{galzerm}, and expanded for large $c$:
\begin{eqnarray}
\nonumber
S[ x]&=&\int_{\mathscr{C}}\text{d}t \, \Omega
 \sqrt{
1-\frac{1}{c^2\Omega^2}a_{ij} \left(v^i -w^i\right)\left(v^j-w^j 
\right)}\\
&=&\int_{\mathscr{C}}\text{d}t\, \Omega\left(1-\frac{1}{2c^2\Omega^2}a_{ij} \left(v^i -w^i\right)\left(v^j-w^j 
\right)+\text{O}\left(\nicefrac{1}{c^4}\right)\right).
\label{rellagzerm}
\end{eqnarray}
Hence, the dynamics \eqref{galact}, disregarding the first term in \eqref{rellagzerm}, which is a Galilean invariant, coincides in the infinite-$c$ limit with the dynamics of the non-relativistic action displayed in \eqref{galactgal}. This shows that \eqref{galzerm} is the natural relativistic spacetime uplift of a Galilean space $\mathscr{S}$ endowed with a Newton--Cartan structure. 

\subsection{\dots to Lewis Carroll}\label{CARdyn}

The Poincar\'e group admits another contraction at vanishing $c$ \cite{Levy}. Although this limit may sound degenerate as particle motion is frozen, it exhibits both an interesting dynamics and a rich mathematical structure. 

A Euclidean space $E_d$ with Cartesian coordinates, accompanied with a real time line $t$ can be equipped with a structure alternative to Newton--Cartan's, known as Carrollian.  This structure is left invariant by the Carrollian group acting as
\begin{equation}
\label{car}
\begin{cases}
t'=t+B_i x^i +t_0,\\
x^{\prime k}=R^{k}_i x^i +x^k_0
\end{cases}
\end{equation}
with all parameters being $(t, \textbf{x})$-independent, and $R^k_i $ the entries of an orthogonal matrix.  

Invariant equations of motion can be considered for extended objects \emph{i.e.} fields rather than particles.  Indeed, at zero velocity of light, a particle cannot move in time but time can define an $\mathbf{x}$-dependent field. The scalar field $t(\mathbf{x})$ describes a $d$-brane, in other words a space-filling object in $E_d$, extended inside a portion of space $\mathscr{V}\subset E_d$.\footnote{Our guide in this section is symmetry, and our goal the adequate Poincar\'e uplift. The precise physical system and the nature of its dynamics are of secondary importance. Other systems with Carrollian symmetry may exist. It is interesting, though, to maintain a dual formulation for the two sides (Galilean and Carrollian), as for objects with dimension-one and codimension-one world-volumes.} Its invariant action 
can be \emph{e.g.}
\begin{equation}
\label{caracteucl}
S[ t]=\int_{\mathscr{V}}\text{d}^dx \mathcal{L}(\pmb{\partial} t)
\end{equation}
with Lagrangian density 
\begin{equation}
\label{carlageucl}
\mathcal{L}(\pmb{\partial} t)=\frac{1}{2}\delta^{ij} \left(\partial_i t-b_i \right)
\left(\partial_j t-b_j\right),
\end{equation}
where 
$b_i$ are constant parameters with inverse-velocity dimension, playing the r\^ole of a constant gauge-field background, and transforming by shift and rotation under \eqref{car}: $b^\prime_i=\left(b_j+ B_j\right) R^{-1j}_{\hphantom{-1}i}$. 

More general Carrollian structures equip Riemannian manifolds $\mathscr{S}$ with metric \eqref{dmet}
and time $t\in \mathbb{R}$. The Carrollian transformations \eqref{car} are realized locally, in the tangent space, and are no longer symmetries. The structure is covariant under \emph{Carrollian diffeomorphisms} 
 \begin{equation}
\label{cardifs} 
t'=t'(t,\textbf{x})\quad \text{and} \quad \textbf{x}^{\prime}=\textbf{x}^{\prime}(\textbf{x})
\end{equation}
with Jacobian functions
\begin{equation}
 \label{carj}
J(t,\mathbf{x})=\frac{\partial t'}{\partial t},\quad j_i(t,\mathbf{x}) = \frac{\partial  t'}{\partial x^{i}},\quad 
J^i_j(\mathbf{x}) = \frac{\partial x^{i\prime}}{\partial x^{j}}.
\end{equation}
The covariant action describing the Carrollian dynamics in the more general case at hand is\footnote{Notice that actions \eqref{caracteucl}, \eqref{caract} and \eqref{relactrp} are all Euclidean-signature (instantonic) because of vanishing $c$.}

\begin{equation}
\label{caract}
S[ t]=\int_{\mathscr{V}\subset\mathscr{S}}\text{d}^dx \sqrt{a} \mathcal{L}(\pmb{\partial} t, t, \mathbf{x}),
\end{equation}
where $a$ stands for the determinant of the matrix $a_{ij}$ and $\mathcal{L}(\pmb{\partial} t, t, \mathbf{x})$ is the Lagrangian density: 
\begin{equation}
\label{carlag}
\mathcal{L}(\pmb{\partial} t, t, \mathbf{x})=\frac{1}{2}a^{ij} \left(\Omega \partial_i t-b_i \right)\left(\Omega \partial_j t-b_j\right).
\end{equation}
Here the components of the metric, the scale factor $\Omega$, and the components of the background gauge field  
$\pmb{b}=b_i \text{d}x^i$ depend all on $( t, \mathbf{x})$.

Under Carrollian diffeomorphisms, the metric transforms as in \eqref{galdifa} \emph{i.e.}
\begin{equation}
\label{cardifa}
a^{\prime ij}=J^i_k J_l^ja^{kl},
\end{equation}
$\Omega$ is rescaled as in \eqref{galdifom} --  where everything now depends both on $t$ and $\mathbf{x}$ -- while the field gradients and the gauge connection obey respectively
\begin{equation}
\label{cardift}
 \partial_k' t'=\left( J \partial_i t+ j_i\right)J^{-1i}_{\hphantom{-1}k},
\end{equation}
and
\begin{equation}
\label{cardifb}
b^{\prime}_{k}=\left( b_i+\frac{\Omega}{J} j_i\right)J^{-1i}_{\hphantom{-1}k}.
\end{equation}
Here 
\begin{equation}
\label{carbeta}
\beta_i=\Omega \partial_i t-b_i
\end{equation}
transform as components of a one-form on $\mathscr{S}$, making the density Lagrangian form-invariant. 

We will now uplift the above structure into a $d+1$-dimensional pseudo-Riemannian manifold $\mathscr{M}$, where the full Poincar\'e group is realized in the tangent space.
Following the pattern used in the Galilean framework, Sec. \ref{GALdyn}, we can recover the general Carrollian structure and its class of diffeomorphisms \eqref{cardifs} in the zero-$c$ limit, starting from a metric on $\mathscr{M}$ of the form:
\begin{equation}
\label{carrp}
\text{d}s^2 =- c^2\left(\Omega \text{d}t-b_i \text{d}x^i
\right)^2+a_{ij} \text{d}x^i \text{d}x^j.
\end{equation}
The form \eqref{carrp} is known as \emph{Randers--Papapetrou}. It is universal, as every metric can be recast in this gauge. Here, it is required for the functions $\Omega(x)$, $a^{ij}(x)$ and $b_i(x)$ to transform as in  \eqref{galdifom}, \eqref{cardifa} and \eqref{cardifb} under a Carrollian diffeomorphism \eqref{cardifs} -- again $x\equiv(x^0=ct,\mathbf{x})$. 
The spacetime Jacobian matrix 
associated with transformations \eqref{cardifs}, reads (using \eqref{carj}):
 \begin{equation}
 \label{carjst} 
J^\mu_\nu(x) = \frac{\partial x^{\mu\prime}}{\partial x^{\nu}}\to
\begin{pmatrix}   J(x) &J_j(x)\\  
0 & J^i_j(\mathbf{x})
 \end{pmatrix} \quad \text{with}\quad J_i=cj_i .
\end{equation}

The Carrollian dynamics captured in the action \eqref{caract} is the zero-$c$ limit of a relativistic instantonic $d$-brane in a spacetime $\mathscr{M}$ with Randers--Papapetrou metric \eqref{carrp}. As already mentioned (footnote \ref{tachin}), in this context instantonic means that the world-volume does not extend in time; it is a kind of codimension-one snap shot materialized in a space-like $d$-dimensional hypersurface $\mathscr{V}$, coordinated with $y^i, i=1,\ldots,d$. Under these assumptions, the Dirac--Born--Infeld action reads:
\begin{equation}
\label{relactrp}
S[h]=\int_{\mathscr{V}}\text{d}^dy \sqrt{h},
\end{equation}
where $h$ is the determinant of the induced metric matrix
\begin{equation}
\label{carh}
h_{ij}=g_{\mu\nu}\frac{\partial x^\mu}{\partial y^i}\frac{\partial x^\nu}{\partial y^j}
\end{equation}
with $g_{\mu\nu}$ the background metric components. 

For the Randers--Papapetrou environment  
displayed in \eqref{carrp}, we find:
\begin{equation}
\label{carhrp}
h_{ij}=\frac{\partial x^k}{\partial y^i}\frac{\partial x^l}{\partial y^j}
\left(a_{kl}-c^2\left(
\Omega \partial_k t-b_k \right)\left(\Omega \partial_l t-b_l\right)
\right).
\end{equation}
In this expression, $ \partial_k t$ stands for $\nicefrac{\partial t}{ \partial x^k}$. Consequently, we implicitly assume that the functions $x^k=x^k(y^i)$ are invertible, which is equivalent to saying that one can choose a gauge where $y^i=x^i$. This is what happens in practice. Indeed,  
one can readily compute the root of the determinant and its expansion in powers of $c^2$. Naming  $\alpha^k_i=\frac{\partial x^k}{\partial y^i}$, we obtain:
\begin{equation}
\label{carsqrthrp}
\sqrt{h}=\det \alpha \sqrt{a}
\left(1-\frac{c^2}{2}a^{kl}\left(
\Omega \partial_k t-b_k \right)\left(\Omega \partial_l t-b_l\right)
+\text{O}\left(c^4\right)\right).
\end{equation}
Hence \eqref{relactrp} becomes 
\begin{equation}
\label{caracrp}
S[h]=\int_{\mathscr{V}}\text{d}^dx
 \sqrt{a}
\left(1-\frac{c^2}{2}a^{kl}\left(
\Omega \partial_k t-b_k \right)\left(\Omega \partial_l t-b_l\right)
+\text{O}\left(c^4\right)\right).
\end{equation}
Neglecting the first term, which is invariant under Carrollian diffeomorphisms \eqref{cardifs}, \eqref{carj}, in the zero-$c$ limit, \eqref{caracrp} describes the same dynamics as \eqref{caract},  \eqref{carlag}. This result, in close analogy with the Galilean discussion in the previous section, shows that the form \eqref{carrp} is well-suited for the zero-$c$ limit.

\section{Fluid dynamics in the non-relativistic limits}

The aim of the present chapter is to exhibit the general fluid equations in the Galilean and Carrollian structures. This is achieved starting from plain relativistic viscous-fluid dynamics in the appropriate background -- Zermelo or Randers--Papapetrou -- and analyzing the associated, infinite or vanishing light-velocity limit. By construction, the equations reached this way are covariant under the corresponding diffeomorphisms. We study here neutral fluids, moving freely \emph{i.e.} subject only to pressure, friction forces and thermal conduction processes. We conclude with some comments on a duality relating the two limits under consideration.

\subsection{Relativistic fluids}\label{recap}

Free relativistic viscous fluids are described in terms of their energy--momentum tensor obeying the set of $d+1$ conservation equations
 \begin{equation}
 \label{conT} 
\nabla_\mu T^{\mu\nu}=0.
\end{equation}
The time component is the energy conservation, the other $d$ spatial ones, momentum conservation, usually called \emph{Euler} equations. 

The energy--momentum tensor is made of a perfect-fluid piece and terms resulting from friction and thermal conduction. It reads:
\begin{equation}\label{T} 
T^{\mu \nu}=(\varepsilon+p) \frac{u^\mu  u^\nu}{c^2} +p  g^{\mu\nu}+   \tau^{\mu \nu}+ \frac{u^\mu  q^\nu}{c^2}   +\frac{u^\nu  q^\mu}{c^2},
\end{equation}
and contains $d+2$ dynamical variables:
\begin{itemize}
\item energy per unit of proper volume (rest density) $\varepsilon$, and pressure $p$;
\item $d$ velocity-field components $u^i$ ($u^0$ is determined by the normalization $\| \text{u} \|^2=-c^2$).
\end{itemize}
 A local-equilibrium thermodynamic equation of state\footnote{We omit here the chemical potential as we assume no independent conserved current.} $p=p(T)$ is therefore needed for completing the system. We also have the usual Gibbs--Duhem relation for the grand potential $-p=\varepsilon-Ts$ with  
  $s=\nicefrac{\partial p}{\partial T}$. The viscous stress tensor $\tau^{\mu \nu}$ and the heat current $q^\mu$ are purely transverse:
\begin{equation}\label{trans}
u^\mu  q_\mu =0, \quad u^\mu    \tau_{\mu \nu}=0, \quad
u^\mu  T_{\mu \nu}=- q_\nu -{\varepsilon} u_\nu, \quad \varepsilon =\tfrac{1}{c^2}T_{\mu \nu} u^\mu u^\nu.
\end{equation}
Hence, they are expressed in terms of $u^i$ and their spatial components $q_i$ and $\tau_{ij}$. 

The quantities $q_i$ and $\tau_{ij}$ capture the physical properties of the out of equilibrium state.  
They are usually expressed as expansions in temperature and velocity derivatives, the coefficients of which characterize the transport phenomena occurring in the fluid. The transport coefficients can be determined either from the underlying microscopic theory, or phenomenologically. In first-order hydrodynamics 
\begin{eqnarray}\label{e1}
&&\tau_{(1)\mu \nu}=-2\eta \sigma_{\mu \nu}-\zeta h_{\mu\nu}\Theta,\\
\label{q1} &&q_{(1)\mu}= -\kappa h_\mu^{\hphantom{\mu}\nu}\left(\partial_\nu T+\frac{T}{c^2}\, a_\nu \right),
\end{eqnarray}
where \footnote{Our conventions for (anti-) symmetrization are 
$A_{(\mu\nu)}=\frac{1}{2}\left(A_{\mu\nu}+A_{\nu\mu}\right)$ and $ 
A_{[\mu\nu]}=\frac{1}{2}\left(A_{\mu\nu}-A_{\nu\mu}\right)$.}
\begin{eqnarray}
&a_\mu =u^\nu \nabla_\nu u_\mu , \quad
\Theta=\nabla_\mu  u^\mu ,& \label{def21}\\
&\sigma_{\mu \nu}= \nabla_{(\mu } u_{\nu )} + \frac{1}{c^2}u_{(\mu } a_{\nu )} -\frac{1}{d} \Theta\,h_{\mu \nu}  ,&
\label{def23}
\\ &\omega_{\mu \nu}= \nabla_{[\mu } u_{\nu ]} +  \frac{1}{c^2}u_{[\mu }a_{\nu] },&\label{def24}
\end{eqnarray}
are the acceleration, the expansion, the shear and the vorticity of the velocity field, with $\eta,\zeta$ the shear and bulk viscosities, and $\kappa$ the thermal conductivity. In the above expressions, 
$h_{\mu \nu} $ is the projector onto the space transverse to the velocity field, and one similarly defines the longitudinal projector $U_{\mu \nu}$:
\begin{equation}
\label{relproj}
h_{\mu\nu}=\dfrac{u_{\mu}u_{\nu}}{c^2}+g_{\mu\nu},\quad U_{\mu\nu}=-\dfrac{u_{\mu}u_{\nu}}{c^2}.
\end{equation}
In three spacetime dimensions, the Hall viscosity appears as well in $\tau_{(1)\mu \nu}$: 
\begin{equation}
\label{relHj}
 -\zeta_{\mathrm{H}}\frac{u^\sigma}{c}\, \eta_{\sigma\lambda(\mu}\, \sigma_{\nu)\rho}\, g^{\lambda\rho},
\end{equation}
with $\eta_{\sigma\lambda\mu}=\sqrt{-g}\, \epsilon _{\sigma\lambda\mu}$.

{ In view of the subsequent steps of our analysis, an important question arises at this stage, which concerns the behaviour of $q_i$ and $\tau_{ij}$ with respect to the velocity of light.} Answering this question requires a microscopic understanding of the fluid \emph{i.e.} a many-body (quantum-field-theory and statistical-mechanics) determination of the transport coefficients. In the absence of this knowledge, we may consider a large-$c$ or small-$c$ expansion of these quantities, in powers of $c^2$ --  irrespective of the derivative expansion. In the same spirit, we could also work out similar expansions for each of the functions entering the metrics \eqref{galzerm} or \eqref{carrp}, as these possibly carry deep relativistic dynamics. The advantage of such an exhaustive analysis would be to set-up general conditions on a relativistic fluid and its spacetime environment for a large-$c$ or a small-$c$ regime to make sense. As a drawback, 
this approach would blur the universality of the equations we want to set. We will therefore adopt a more pragmatic attitude and  assume that $\Omega$, $b_i$, $w^j$ and $a_{ij}$ are $c$-independent. 
Regarding the viscous stress tensor $\tau_{ij}$, we will assume the following behaviours:
\begin{equation}
\label{sigexpG}
\tau_{ij}=-\Sigma^{\text{G}}_{\hphantom{\text{G}}ij}
\end{equation}
or
\begin{equation}
\label{sigexpC}
\tau^{ij}=-\frac{\Sigma^{\text{C}ij}}{c^2}-\Xi^{ij}.
\end{equation}
The first is appropriate for the Galilean limit. It is standard and  considered \emph{e.g.} in \cite{Landau}, where  
$\Sigma^{\text{G}}_{\hphantom{\text{G}}ij}$ is named  $\sigma^{\prime}_{ij}$. For the Carrollian dynamics, our choice is inspired by flat-spacetime holography (see \cite{CMPPS2}). Similarly, for the heat current, we will adopt
\begin{eqnarray}
\label{QexpG}
q_i&=&Q^{\text{G}}_{\hphantom{\text{G}}i},
\\
\label{QexpC}
q^i&=&Q^{\text{C}i}+{c^2}\pi^i,
\end{eqnarray}
in Galilean and Carrollian dynamics, respectively. Although kinematically poorer -- because at rest, Carrollian fluids carry a richer internal information than their Galilean pendants since both the heat current and the viscous tensor are doubled in the above ansatz. Observe the position of the spatial indices, different for the two cases under consideration. They are designed to be covariant under different classes of diffeomorphisms. 

One should finally notice that, in writing the energy--momentum tensor  \eqref{T}, we have not made any assumption regarding the hydrodynamic frame, which is therefore left 
generic.\footnote{The freedom of choosing the hydrodynamic frame was raised in  \cite{Landau}. Modern discussions can be found in \cite{RZ, Romatschke:2009im, Kovtun:2012rj} (see also \cite{Ciambelli:2017wou}).}
 There are two reasons for this. The first is the absence of a conserved relativistic current, which makes hydrodynamic-frame conditions delicate. Further subtleties arise when studying the system in special limits such as the Galilean, where the relativistic arbitrariness for the velocity field is lost, due to the decoupling of mass and energy. This is the second reason.

\subsection{Galilean fluid dynamics from Zermelo background} \label{FLUIDS}

\subsubsection*{The essence of the classical limit}

We will consider in the following the ordinary non-relativistic limit of fluid equations, formally reached at infinite $c$. { The physical validity of this situation is based on two assumptions. 

The first is kinematical: it presumes that the global velocity of the fluid with respect to the observer is small compared to $c$. This 
is easily implemented using the Zermelo form of the metric \eqref{galzerm}, where the control parameter for the validity of the classical limit is $\left\vert\frac{\mathbf{v}-\mathbf{w}}{c} \right\vert$. We find
\begin{equation}
\label{galvel}
\begin{cases}
u^0=\gamma c=\dfrac{c}{\Omega}+\text{O}\left(\nicefrac{1}{c}\right),
\quad
u_0=-c\Omega+\text{O}\left(\nicefrac{1}{c}\right),
\\
u^i=\gamma v^i=\dfrac{v^i}{\Omega}+\text{O}\left(\nicefrac{1}{c^2}\right),
\quad u_i=\dfrac{v_i-w_i}{\Omega}+\text{O}\left(\nicefrac{1}{c^2}\right).
\end{cases}
\end{equation}

The second is microscopic. The internal particle motion should also be Galilean,  in other words the energy density should be large compared to the pressure: $\varepsilon\gg p$. This sets restrictions on the equation of state, as not every equation of state is compatible with such a microscopic assumption.\footnote{For example, the conformal equation of state, $\varepsilon=dp $ is not compatible with the non-relativistic limit at hand.}}

An important consequence of the microscopic assumption is the separation of mass and energy, now both independently conserved. It is customary to introduce the following:
\begin{itemize}
\item $\varrho$ the usual mass per unit of volume (mass density);
\item $\varrho_0$ the usual mass per unit of proper volume (rest-mass density);
\item $e$ the internal energy per unit of  mass;
\item $h$ the enthalpy  per unit of  mass.
\end{itemize}
These local thermodynamic quantities are related as
\begin{equation}
\label{galdef}
\begin{cases}
\varepsilon=\left(e+c^2\right) \varrho_0,\\
h=e+\frac{p}{\varrho},\\
\varrho_0=\dfrac{\varrho}{\Omega \gamma}=\varrho \sqrt{1-\left(\frac{\mathbf{v}-\mathbf{w}}{c\Omega}\right)^2}\approx\varrho-\frac{\varrho}{2}\left(\frac{\mathbf{v}-\mathbf{w}}{c\Omega}\right)^2,
\end{cases}
\end{equation}
where we have used Eq. \eqref{gamzerm} for the Lorentz factor $\gamma$, and expanded it for small $\left\vert\frac{\mathbf{v}-\mathbf{w}}{c} \right\vert$.

\subsubsection*{The structure of the equations}

The fluid equations are the conservation \eqref{conT} of the energy--momentum tensor \eqref{T}, in the background \eqref{galzerm}. It is computationally wise to split these equations as:
 \begin{equation}
 \label{conTgal} 
\nabla_\mu T^{\mu0}=0,\quad \nabla_\mu T^{\mu}_{\hphantom{\mu}i}=0.
\end{equation}
Indeed, applying a Galilean diffeomorphism \eqref{galdifs},  \eqref{galjst}, the time components up and space components down transform faithfully and irreducibly. On the divergence of the energy--momentum tensor we find:
 \begin{equation}
 \label{conTgaljst} 
\nabla_\mu^{\prime} T^{\prime\mu0}=J \nabla_\mu T^{\mu0},\quad \nabla_\mu^{\prime} T^{\prime\mu}_{\hphantom{\prime\mu}i}=J^{-1l}_{\hphantom{-1}i}  \nabla_\mu T^{\mu}_{\hphantom{\mu}l}.
\end{equation}
Hence, the two sets of equations \eqref{conTgal} do not mix\footnote{They do mix for general diffeomorphisms though.} and have furthermore a $d$-dimensional covariant transformation, which is our goal for the Galilean fluid dynamics.

The expressions displayed so far are fully relativistic. The next step is to consider the large-$c$ regime. In this regime, Eqs. \eqref{conTgal} can be expanded in powers of $\nicefrac{1}{c}$.
This expansion must be performed with care as the time equation needs an extra $c$  factor with respect to the next $d$ spatial equations because it describes the evolution of energy, which is a momentum multiplied by $c$. We find:\footnote{Had we considered $\Omega=\Omega(t,\mathbf{x})$, the divergence  $\nabla_\mu T^{\mu}_{\hphantom{\mu}i}$ would have exhibited an extra, dominant  term in the large-$c$ limit: $c^2 \partial_i\ln \Omega$. The spatial conservation equation,  $\nabla_\mu T^{\mu}_{\hphantom{\mu}i}=0$, would then automatically require the $\mathbf{x}$-independence for $\Omega$.
Notice also the rescaling  by $\Omega$ in \eqref{conTgalexp0},  which guarantees that $\mathcal{C}$ and $\mathcal{E}$ are invariants under Galilean diffeomorphisms, see \eqref{galCEdif}.}
 \begin{eqnarray}
 \label{conTgalexp0} 
c\nabla_\mu T^{\mu0}&=& c^2\frac{\mathcal{C}}{\Omega}+\frac{\mathcal{E}}{\Omega}+ \text{O}\left(\frac{1}{c^2}\right),
\\
\label{conTgalexpi} 
\nabla_\mu T^{\mu}_{\hphantom{\mu}i}&=&\mathcal{M}_i+ \text{O}\left(\frac{1}{c^2}\right).
\end{eqnarray}
At infinite $c$ this leads to $d+2$ equations (rather than $d+1$, since in the Galilean limit, mass and energy are separately conserved) for $\varrho$, $e$, $p$ and $v^i$:
\begin{itemize}
\item continuity equation (mass conservation) $\mathcal{C}=0$;
\item energy conservation $\mathcal{E}=0$;
\item momentum conservation $\mathcal{M}_i=0$;
\end{itemize}
this system is completed with the equation of state $p=p(e,\varrho)$.

It is important to stress that Galilean diffeomorphisms  \eqref{galdifs},  \eqref{galj} do not involve $c$, and consequently they do not mix the various terms in the expansions \eqref{conTgalexp0}   and \eqref{conTgalexpi}. All $d+2$ fluid equations reached this way on general backgrounds\footnote{We stress again that here, as for instance in \cite{Jensen:2014ama,Geracie:2015xfa}, Galilean fluids evolve on general, curved and time-dependent spaces $\mathscr{S}$, as opposed to other works on non-relativistic fluid dynamics (see \emph{e.g.} \cite{Banerjee:2014mka}).} are guaranteed to be covariant under Galilean diffeomorphisms, and this was one motivation of our work.

\subsubsection*{The dissipative tensors in Zermelo background}

Before displaying the advertised equations, we would like to elaborate on the two tensors which capture the deviation of the real fluid with respect to the perfect one: the heat current and the viscous stress tensor. 

Orthogonality conditions \eqref{trans} allow to express every component of these tensors in terms of $q_i$ and $\tau_{ij}$. We assume here the Zermelo form of the metric \eqref{galzerm}, and a fluid velocity field as in \eqref{vel}, \eqref{gamzerm}.  We find
\begin{equation}
 \label{galheat} 
q_0=-\frac{v^i q_i }{c},\quad q^0=\frac{\left(v^i-w^i\right)q_i}{c\Omega^2},\quad
q^i=a^{ij}q_j
+\frac{w^i\left(v^j-w^j\right)q_j}{c^2\Omega^2}.
\end{equation}
Similarly, the components of the stress tensor are obtained from the $\tau_{ij}$s. For example:
\begin{equation}
 \label{galfs} 
\tau_{00}= \frac{v^kv^l \tau_{kl}}{c^2},\quad \tau_{0j}=- \frac{v^k \tau_{kj}}{c},\quad \tau^0_{\hphantom{0}j}=- \frac{\left(v^k-w^k\right)\tau_{kj}}{c\Omega^2}, \quad 
\tau^{00}=\frac{\left(v^k-w^k\right)\left(v^l-w^l\right)\tau_{kl}}{c^2\Omega^4},\ldots
\end{equation}

We now define $Q^{\text{G}}_{\hphantom{\text{G}}i}=q_i$ as anticipated in \eqref{QexpG}, and 
\begin{equation}
 \label{galheatd} 
Q^{\text{G}i}= a^{ij}Q^{\text{G}}_{\hphantom{\text{G}}j}.
\end{equation}
Similarly, calling for $\Sigma^{\text{G}}_{\hphantom{\text{G}}ij}$ introduced in
\eqref{sigexpG}, we define 
\begin{equation}
 \label{galfsd} 
\Sigma^{\text{G}\hphantom{i}j}_{\hphantom{\text{G}}i}= \Sigma^{\text{G}}_{\hphantom{\text{G}}ik}a^{kj},\quad
\Sigma^{\text{G}ij} = a^{ik}\Sigma^{\text{G}\hphantom{k}j}_{\hphantom{\text{G}}k}.
\end{equation}
Using the generic transformation rules of $q_\mu$ and $\tau_{\mu\nu}$ under spacetime diffeomorphisms, we find that $\pmb{Q}^{\text{G}}$ and $\pmb{\Sigma}^{\text{G}}$ introduced above, appearing as classical $c$-independent objects,  transform as they should, namely as $d$-dimensional tensors under Galilean diffeomorphisms  \eqref{galdifs},  \eqref{galjst}:
\begin{eqnarray}
 \label{galheatddiff} 
&Q^{\text{G}\prime}_{\hphantom{\text{G}\prime}i} = Q^{\text{G}}_{\hphantom{\text{G}}k} J^{-1k}_{\hphantom{-1}i}, \quad
\quad
Q^{\text{G}\prime i}=J^{i}_{k}Q^{\text{G}k},&
\\
 \label{galfsddiff} 
& \Sigma^{\text{G}\prime }_{\hphantom{\text{G}\prime }ij} 
=J^{-1k}_{\hphantom{-1}i}J^{-1l}_{\hphantom{-1}j}\Sigma^{\text{G}}_{\hphantom{\text{G}}kl}, \quad
\quad
\Sigma^{\text{G}\prime\hphantom{i}j}_{\hphantom{\text{G}\prime}i}
=J^{-1k}_{\hphantom{-1}i} \Sigma^{\text{G}\hphantom{k}l}_{\hphantom{\text{G}}k} J^{j}_{l}, \quad
\Sigma^{\text{G}\prime  ij} =\Sigma^{\text{G} kl} J^{i}_{k}J^{j}_{l}.&
\end{eqnarray}

\subsubsection*{Continuity and energy conservation}

Using Eq. \eqref{T} for the energy--momentum tensor $T^{\mu\nu}$ with $g^{\mu\nu}$  and $u^\mu$ given in  \eqref{galzerm} and  \eqref{vel}, using Eqs. \eqref{galheat}, \eqref{galheatd} for the heat current and  \eqref{galfs}, \eqref{galfsd} for the stress tensor as well as the definitions \eqref{galdef}, we can perform the large-$c$ expansion of the relativistic energy conservation equation \eqref{conTgalexp0}. This requires the expansion of the Christoffel symbols, displayed in App.  \ref{zerapp}. 

We find the following at $\text{O}(c^2)$:
\begin{equation}
\label{galC} 
\mathcal{C}=\frac{\partial_t \sqrt{a}\varrho }{\Omega\sqrt{a}}+\frac{1}{\Omega} \nabla_i \varrho v^i,
\end{equation}
where $a$ stands for the determinant of the $d$-dimensional metric $a_{ij}(t,\mathbf{x})$, and $\nabla_i$ is the Levi--Civita covariant derivative associated with $a_{ij}(t,\mathbf{x})$ and Christoffel symbols given in \eqref{dgamma}.
The standard continuity equation $\mathcal{C}=0$ is thus recovered.
It is customary to decompose $\mathcal{C}$ in \eqref{galC} as 
\begin{equation}
\label{galCpart} 
\frac{\partial_t \sqrt{a}\varrho }{\Omega\sqrt{a}}+\frac{1}{\Omega}\nabla_i \varrho v^i= \frac{1}{\Omega}\frac{\text{d}\varrho}{\text{d}t}
+\varrho \theta^{\text{G}},
\end{equation}
where
\begin{equation}
\label{galfder} 
\frac{\text{d}}{\text{d}t}= \partial_t +v^i\nabla_i
\end{equation}
is the \emph{material derivative}, and 
\begin{equation}
\label{galexp} 
\theta^{\text{G}} = \frac{1}{\Omega}\left(
\partial_t \ln \sqrt{a}+\nabla_i v^i\right)
\end{equation}
the \emph{effective Galilean fluid expansion}. The latter combines the divergence of the fluid congruence with the logarithmic expansion of the volume form  to produce a genuine scalar under Galilean diffeomorphisms \eqref{galdifs},  \eqref{galj} (see Eqs. \eqref{galdifom} and \eqref{gal1}). The material derivative \eqref{galfder}, in the form $\frac{1}{\Omega}\frac{\text{d}}{\text{d}t}$, is also an ``invariant'' when acting on a scalar function. This is due to \eqref{galdifv}, \eqref{galdelt} and \eqref{galdelj}. When acting on arbitrary tensors, it should be supplemented with the appropriate $\mathbf{w}$-connection terms, as shown in the appendix, Eq. \eqref{galfcovder}. 

At the next $\text{O}(c^0)$ order, we obtain:
\begin{eqnarray}
\mathcal{E}&=& \frac{1}{\Omega\sqrt{a}}\partial_t\left(
\sqrt{a}\varrho\left( e+\frac{1}{2}\left(\frac{\mathbf{v}-\mathbf{w}}{\Omega} \right)^2\right)
\right)
+\frac{1}{\Omega}\nabla_i \left(\varrho v^i \left( e+\frac{1}{2}\left(\frac{\mathbf{v}-\mathbf{w}}{\Omega} \right)^2\right)
\right)\nonumber \\
&&+\frac{1}{\Omega}\nabla_i\left(\left(v^j-w^j\right)\left(p\delta^i_j-\Sigma^{\text{G}\hphantom{j}i}_{\hphantom{\text{G}}j}  \right)\right)+ \nabla_iQ^{\text{G}i} +\frac{1}{\Omega}\Pi^{\text{G}ij}
\left(\nabla_ iw_j+\frac{1}{2}\partial_t a_{ij}\right)
\label{galE} 
\\&=&\dfrac{\varrho}{\Omega}\frac{\text{d}}{\text{d}t}\left( e+\frac{1}{2}\left(\frac{\mathbf{v}-\mathbf{w}}{\Omega} \right)^2\right)
+\frac{1}{\Omega}\nabla_i\left(p\left(v^i-w^i\right)\right)+ \nabla_iQ^{\text{G}i} \nonumber\\
&&-\frac{1}{\Omega}\nabla_i\left(\left(v^j-w^j\right)\Sigma^{\text{G}\hphantom{j}i}_{\hphantom{\text{G}}j} \right)+\frac{1}{\Omega}\Pi^{\text{G}ij}
\left(\nabla_ iw_j+\frac{1}{2}\partial_t a_{ij}\right),
\label{galEbis} 
\end{eqnarray}
where the alternative expression \eqref{galEbis} is obtained from  \eqref{galE} using the continuity equation $\mathcal{C}=0$.  Here we introduced
\begin{equation}
\Pi^{\text{G}ij}
=
\varrho  \frac{\left(v^i-w^i\right)\left(v^j-w^j\right)}{\Omega^2}
+p a^{ij}-\Sigma^{\text{G}ij} ,
\label{galPI} 
\end{equation}
the components of the Galilean energy--momentum tensor, following \cite{Landau}. They are expressed in terms of the fluid velocity, measured in an inertial-like frame, \emph{i.e.} $\mathbf{v}-\mathbf{w}$, and transform under Galilean diffeomorphisms \eqref{galdifs},  \eqref{galj} as a genuine rank-two $d$-dimensional  tensor on $\mathscr{S}$ (one uses 
\eqref{galdifa},
\eqref{galdifv},
\eqref{galdifw},
\eqref{galdifom},
and
\eqref{galfsddiff}):
\begin{equation}
\Pi^{\text{G} ij\prime} =J_{k}^{i} J_{l}^{j}\Pi^{\text{G}kl} .
\end{equation}

Equation $\mathcal{E}=0$ is the Galilean energy conservation equation for a viscous fluid in motion on arbitrary, time-dependent $d$-dimensional space $\mathscr{S}$, and observed from an arbitrary frame (moving at velocity $-\mathbf{w}(t,\mathbf{x})$ with respect to a local inertial frame). 
In a short while, we will recast this equation in a suitable form for recognizing the underlying phenomena.
Notice that both friction and thermal conduction occur, driven by the viscous stress tensor $\pmb{\Sigma}^{\text{G}}$ and the heat current $\pmb{Q}^{\text{G}}$.  As opposed to the energy-conservation equation at hand,
the continuity (mass-conservation) equation depends neither on the motion of the observer ($\mathbf{w}$) nor on the friction properties of the fluid. This is expected because energy  is frame-dependent while mass it is not.

One can check that under Galilean diffeomorphisms   \eqref{galdifs},  \eqref{galj}:
\begin{equation}
\label{galCEdif} 
\mathcal{C}^\prime= \mathcal{C},\quad \mathcal{E}^\prime= \mathcal{E}.
\end{equation}
In order to show this, it is convenient to recognize some well-behaved blocks in the expressions at hand, based on the quoted transformation rules. We have gathered this information in App. \ref{zerapp}, Eqs. \eqref{gal-1}--\eqref{gal3}. For \eqref{galCEdif}, we also need \eqref{galheatddiff}, \eqref{galfsddiff}.

\subsubsection*{Euler equation}

Following the same pattern, we can process the large-$c$ behaviour of the relativistic momentum-conservation equations. Along with \eqref{conTgalexpi} we find:
\begin{equation}
\mathcal{M}_i= \frac{1}{\Omega\sqrt{a}}\partial_t\left(
\sqrt{a}\varrho \frac{v_i-w_i}{\Omega}\right)
+\dfrac{1}{\Omega}\nabla_j \left(\varrho w^j  \left(\frac{v_i-w_i}{\Omega}\right)
\right)
+\frac{\varrho}{\Omega}
\left(\dfrac{v^j-w^j}{\Omega}\right)\nabla_ iw_j
+\nabla_j \Pi^{\text{G}\hphantom{i}j}_{\hphantom{\text{G}}i} 
\label{galM} 
\end{equation}
with $\Pi^{\text{G}\hphantom{i}j}_{\hphantom{\text{G}}i} $ as in \eqref{galPI}. 
The equation $\mathcal{M}_i=0$ is the ultimate  generalization of the standard Euler equation,  displayed \emph{e.g.} in Ref. \cite{Landau}. It is remarkably simple.
The second and third terms in \eqref{galM} contribute to inertial forces (Coriolis, centrifugal etc.), and are usually absent in Euclidean space with inertial frames. Together with the first term, they provide the components of a one-form on $\mathscr{S}$ transforming as $\frac{\mathbf{v}-\mathbf{w}}{\Omega}$ (see \eqref{gal4}, \eqref{gal5}). This is also how $\mathcal{M}_i$ behave under Galilean diffeomorphisms   \eqref{galdifs}, \eqref{galj}:
\begin{equation}
\label{galMdif} 
\mathcal{M}^\prime_i = J^{-1l}_{\hphantom{-1}i} \mathcal{M}_l.
\end{equation}

The Euler equation \eqref{galM} contains the \emph{acceleration} $\pmb{\gamma}^{\text{G}}=\gamma^{\text{G}}_{\hphantom{G}i}\text{d}x^i$ of the Galilean fluid. This is defined covariantly as 
\begin{equation}
\label{relalim}
a_i = \gamma^{\text{G}}_{\hphantom{C}i} +\text{O}\left(\nicefrac{1}{c^2}\right)
\end{equation}
with $a_i$ the spatial components of the relativistic fluid acceleration as in \eqref{def21}. We find:
 \begin{equation}
\label{galg} 
\Omega^2\gamma^{\text{G}}_{\hphantom{G}i} = \Omega \frac{\text{d}\nicefrac{v_i}{\Omega}}{\text{d}t}-\Omega\partial_t\nicefrac{w_i}{\Omega}-\frac{1}{2}\partial_i \mathbf{w}^2 -v^j\left(\partial_j w_i-\partial_iw_j\right)
\end{equation}
with $\nicefrac{\text{d}}{\text{d}t}$ defined in \eqref{galfder}. In this expression, $\gamma^{\text{G}}_{\hphantom{G}i} $ appear as the components of the acceleration  in the local inertial frame and
$ \frac{\text{d}\nicefrac{v_i}{\Omega}}{\Omega\text{d}t}$ are
the components of the effectively measured acceleration in the coordinate frame at hand. In the right hand side, the second term is the dragging acceleration, the third accounts for the centrifugal acceleration, and the last is Coriolis contribution.  We can alternatively write \eqref{galg} as
 \begin{equation}
\label{galgal} 
\gamma^{\text{G}}_{\hphantom{G}i}= \frac{\text{d}\nicefrac{(v_i-w_i)}{\Omega}}{\Omega\text{d}t}-\frac{1}{2}\partial_i \frac{\mathbf{w}^2}{\Omega^2} +\frac{v^j}{\Omega}\nabla_i\frac{w_j}{\Omega}=\frac{\text{D}\nicefrac{(v_i-w_i)}{\Omega}}{\Omega\text{d}t},
\end{equation}
where we used the \emph{Galilean covariant time-derivative} \eqref{galfcovderf} in the second equality.

By construction, the $\gamma^{\text{G}}_{\hphantom{G}i} $s transform as components of a genuine $d$-dimensional form and $\gamma^{\text{G}i}=a^{ij}\gamma^{\text{G}}_{\hphantom{G}j}$  as a vector, under Galilean diffeomorphisms   \eqref{galdifs}, \eqref{galj}:
\begin{equation}
\label{galgdif} 
\gamma^{\text{G}\prime}_{\hphantom{G}i} = J^{-1l}_{\hphantom{-1}i} \gamma^{\text{G}}_{\hphantom{G}l}.
\end{equation}
One can also check explicitly the covariance of \eqref{galg} using \eqref{gal5}.
Using $\gamma^{\text{G}}_{\hphantom{G}i}$ in \eqref{galg} and the expression \eqref{galPI} for the Galilean energy--momentum tensor, we can recast $\mathcal{M}_i$ in \eqref{galM} \emph{\`a la} Euler:  
\begin{equation}
\mathcal{M}_i= \varrho \gamma^{\text{G}}_{\hphantom{G}i} +\partial_i p -\nabla_j  \Sigma^{\text{G}\hphantom{i}j}_{\hphantom{\text{G}}i}.
\label{galMEuler} 
\end{equation}

\subsubsection*{Energy and entropy}

The momentum equation $\mathcal{M}_i=0$ together with continuity equation $\mathcal{C}=0$ can also be used in order to provide a sharper expression for $\mathcal{E}$ given in \eqref{galE}, and leading to:
\begin{equation}
\frac{1}{\Omega\sqrt{a}}\partial_t\left(
\sqrt{a}\varrho\left( e+\frac{\mathbf{v}^2-\mathbf{w}^2}{2\Omega^2}\right)
\right)=
-\nabla_i \Pi^{\text{G}i} -\frac{1}{2\Omega}\Pi^{\text{G}ij}
\partial_t a_{ij}+\varrho\frac{v_j-w_j}{\Omega^2}\partial_t\frac{w^j}{\Omega}.
\label{galEEuler} 
\end{equation}
In this equation, $\varrho\left( e+\frac{\mathbf{v}^2-\mathbf{w}^2}{2\Omega^2}\right)$ is the total energy density of the fluid in the natural, non-inertial frame. The energy density has three contributions: $e\varrho$ as internal energy, the kinetic energy $\nicefrac{\varrho\mathbf{v}^2}{2\Omega^2}$, and the potential energy of inertial forces $\nicefrac{-\varrho\mathbf{w}^2}{2\Omega^2}$ (see \eqref{galen} for the free particle paradigm). Furthermore
\begin{equation}
\Pi^{\text{G}i}=\varrho \frac{v^i}{\Omega} \left( h+\frac{\mathbf{v}^2-\mathbf{w}^2}{2\Omega^2} \right)+ Q^{\text{G}i}
-\frac{v^j}{\Omega} \Sigma^{\text{G}\hphantom{j}i}_{\hphantom{\text{G}}j} 
\label{galMEnflux} 
\end{equation}
appears as the \emph{Galilean energy flux}. It receives contributions from the enthalpy, the kinetic and inertial-potential energies, as well as from dissipative processes: thermal conduction and friction, with the corresponding heat current $\pmb{Q}^{\text{G}}$ and viscous stress current $\nicefrac{-\mathbf{v}\cdot \pmb{\Sigma}^{\text{G}}}{\Omega}$. The general energy conservation equation $\mathcal{E}=0$ has now a simple interpretation: the time variation of energy in a local domain is due to the energy flux through the frontier plus the external work due to the time dependence of $a_{ij}$ and $w^i$ (as for the free particle \eqref{galenth}).

Dissipative processes create entropy. One can readily determine the variation of the latter by recasting the energy variation in a manner slightly different than \eqref{galEEuler}. For that we compute $\mathcal{E}-\frac{v^i-w^i}{\Omega}\mathcal{M}_i$ with \eqref{galE}, \eqref{galgal}, \eqref{galMEuler}. We find, using continuity and \eqref{galexp}:
\begin{equation}
\mathcal{E}-\frac{v^i-w^i}{\Omega}\mathcal{M}_i=
\frac{\varrho}{\Omega}\frac{\text{d}e}{\text{d}t} +p\theta^{\text{G}}+\nabla_i Q^{\text{G}i}-\frac{1}{\Omega} \Sigma^{\text{G}ij} 
\left(\nabla_ iv_j+\frac{1}{2}\partial_t a_{ij}\right).
\label{galEEuler2} 
\end{equation}
In this expression, we can trade the energy per mass $e$, for the entropy per mass $s$, obeying 
\begin{equation}
\text{d}e=T\text{d}s-p\text{d}v=T\text{d}s+\frac{p}{\varrho^2}\text{d}\varrho,
\label{galthermo} 
\end{equation}
where $v=\nicefrac{1}{\varrho}$.
Substituting \eqref{galthermo} in \eqref{galEEuler2}, and trading $\nicefrac{\text{d}\varrho}{\text{d}t}$ for $-\Omega \varrho \theta^{\text{G}}$ (continuity), we obtain finally, owing to $\mathcal{E}= \mathcal{M}_i=0$:
\begin{equation}
\frac{\varrho T}{\Omega}\frac{\text{d}s}{\text{d}t} =\frac{1}{\Omega} \Sigma^{\text{G}ij} 
\left(\nabla_ iv_j+\frac{1}{2}\partial_t a_{ij}\right)-\nabla_i Q^{\text{G}i}.
\label{galentropy} 
\end{equation}
The entropy is not conserved as a consequence of friction and heat conduction, which encode dissipative processes. { The latter are globally captured in a \emph{generalized dissipation function}
\begin{equation}
\psi=\frac{1}{\Omega} \Sigma^{\text{G}ij} 
\left(\nabla_ iv_j+\frac{1}{2}\partial_t a_{ij}\right)-\nabla_i Q^{\text{G}i},
\label{galdisfun} 
\end{equation}
appearing both in energy and entropy equations \eqref{galEEuler2},
 \eqref{galentropy}. Observe that $\psi$ depends explicitly on Christoffel symbols as well as on the time variation of the metric. Hence time dependence and inertial forces contribute the dissipation phenomena.\footnote{  The effect of inertial forces on dissipation has been recently studied by simulation of flows on curved static films without heat current (\emph{i.e.} $d=2$, $\Omega = 1$, $\mathbf{w}=0$, $\partial_t a_{ij}=0$, $\pmb{Q}^{\text{G}}=0$) \cite{Debus}. One might consider performing similar simulations or experiments for probing the more general sources of dissipation present in \eqref{galdisfun}.}}

\subsubsection*{First-order Galilean hydrodynamics and incompressible fluids}

The viscous stress tensor  $\pmb{\Sigma}^{\text{G}}$ and the heat current $\pmb{Q}^{\text{G}}$ are constructed phenomenologically as velocity and temperature derivative expansions. Since these objects transform tensorially under Galilean diffeomorphisms (see \eqref{galheatddiff}, \eqref{galfsddiff}), they must be expressed in terms of tensorial derivative quantities. 

At first order, we have $\theta^{\text{G}}$ defined in \eqref{galexp}, which is an invariant, and 
\begin{equation}
\label{galsymdv} 
\frac{1}{\Omega}\left(\nabla_{(k}v_{l)}+\frac{1}{2}\partial_ta_{kl} \right),
\end{equation}
which is a rank-two symmetric tensor (see \eqref{gal3}). We can therefore set  
\begin{eqnarray}
\label{galsigG1} 
&\Sigma^{\text{G}}_{(1)ij}= 2\eta^{\text{G}}\xi^{\text{G}}_{\hphantom{\text{G}}ij}
+ \zeta^{\text{G}} a_{ij}\theta^{\text{G}},
&
\\
\label{galQG1} 
&Q^{\text{G}}_{(1)i}=-\kappa^{\text{G}} \partial_i T.&
\end{eqnarray}
The transport coefficients are as usual the shear viscosity $\eta^{\text{G}}$, coupled to the \emph{Galilean shear}, 
\begin{equation}
\label{galshear} 
\xi^{\text{G}}_{\hphantom{\text{G}}ij}=\dfrac{1}{\Omega}\left(\nabla_{(i}v_{j)}+\frac{1}{2}\partial_ta_{ij} \right)
-\dfrac{1}{d} a_{ij}\theta^{\text{G}},
\end{equation}
which receives also contributions from the derivative of the metric; the bulk viscosity  $\zeta^{\text{G}}$, coupled to the Galilean expansion, and the thermal conductivity $\kappa^{\text{G}}$ coupled to the temperature gradient. 

Using the definitions of relativistic  expansion and shear \eqref{def21}, \eqref{def23}, we can find their behaviour at large $c$ in the Zermelo background:
\begin{eqnarray}
\label{galsiglim} 
&\sigma_{ij}=\xi^{\text{G}}_{\hphantom{\text{G}}ij}+
\text{O}\left(\nicefrac{1}{c^2}\right),
&
\\
\label{galexplim} \hphantom{\text{G}}
&\Theta=\theta^{\text{G}} +
\text{O}\left(\nicefrac{1}{c^2}\right).&
\end{eqnarray}
For completeness we also display the leading behaviour of the vorticity \eqref{def24}, even though it plays no r\^ole in first-order hydrodynamics:
\begin{equation}
\label{galomlim}
\omega_{ij}=\frac{1}{\Omega}\left(\partial_{[i}(v-w)_{j]}\right)+
\text{O}\left(\nicefrac{1}{c^2}\right).
\end{equation}
Since furthermore the transverse projector \eqref{relproj} is $h_{ij}=a_{ij} +
\text{O}\left(\nicefrac{1}{c^2}\right)$, using \eqref{e1} and \eqref{q1} together with \eqref{sigexpG}
and \eqref{relalim}, we find indeed \eqref{galsigG1} and \eqref{galQG1} (by definition $Q^{\text{G}}_{\hphantom{\text{G}}i}=q_i$). It is important to stress at this point that transport coefficients are determined as modes of microscopic correlation functions, and are therefore sensitive to the velocity of light. In writing \eqref{sigexpG}, we have assumed the following large-$c$ behaviour: 
\begin{equation}
\label{relgaltr}
\eta=\eta^{\text{G}}+\text{O}\left(\nicefrac{1}{c^2}\right),\quad
\zeta=\zeta^{\text{G}}+\text{O}\left(\nicefrac{1}{c^2}\right),\quad 
\kappa=\kappa^{\text{G}}+\text{O}\left(\nicefrac{1}{c^2}\right).
\end{equation}

The case $d=2$ is peculiar because $\Sigma^{\text{G}}_{(1)ij}$ admits an extra term:
\begin{equation}
\label{galHj}
 \zeta^{\text{G}}_{\mathrm{H}}\,  \eta^{\vphantom{\text{G}}}_{k(i}\, \xi^{\text{G}}_{\hphantom{\text{G}}j)l}\, a^{kl}= \frac{ \zeta^{\text{G}}_{\mathrm{H}}}{2\Omega}
 \left(\eta_{k(i}\, \nabla_{j)}v^k+\eta_{k(i}\, a_{j)l}\left(
\nabla^kv^l 
-\frac{\partial_t
\sqrt{a}a^{kl}
}{\sqrt{a}}-a^{kl}\nabla_mv^m
 \right)
 \right)
\end{equation}
with $\eta_{kl}=\sqrt{a}\, \epsilon _{kl}$. This is indeed (up to a global sign) the infinite-$c$ limit of the relativistic Hall-viscosity contribution in three spacetime dimensions given in \eqref{relHj}, assuming again $\zeta_{\mathrm{H}}=\zeta^{\text{G}}_{\mathrm{H}}+\text{O}\left(\nicefrac{1}{c^2}\right)$.

We can now combine the first-derivative contribution  \eqref{galsigG1} of the viscous stress tensor with expression \eqref{galMEuler}  for $\mathcal{M}_i$ in order to obtain the momentum conservation equation $\mathcal{M}_i=0$ of first-order Galilean hydrodynamics. We obtain
\begin{equation}
\label{galMEuler1} 
\varrho \gamma^{\text{G}}_{\hphantom{G}i} +\partial_i p -\frac{\eta^{\text{G}}}{\Omega}\left(\Delta v_i +r_ i^{\hphantom{i}j}v_j
+a_{ik} a^{jl}\partial_t\gamma^k_{jl}\right)-\left(\zeta^{\text{G}}+\frac{d-2}{d}\eta^{\text{G}}
\right)\partial_i \theta^{\text{G}}
=0,
\end{equation}
where $\Delta=\nabla^i\nabla_ i$ is the Laplacian operator in $d$ dimensions and $r_ {ij}$ the Ricci tensor of the $d$-dimensional  Levi--Civita connection $\gamma^k_{ij}$.  Similarly, substituting 
 \eqref{galsigG1},  \eqref{galQG1} and  \eqref{galshear} in \eqref{galentropy}, we find the entropy equation in first-order hydrodynamics on general backgrounds:
 \begin{equation}
\frac{\varrho T}{\Omega}\frac{\text{d}s}{\text{d}t} =\frac{2\eta^{\text{G}}}{\Omega^2} 
\left(\big(\nabla^ iv^j\big)\left(\nabla_ iv_j\right)+\big(\nabla^ iv^j\big)\partial_t a_{ij}
-\frac{1}{4}\big(\partial_t a^{ij}\big)\big(\partial_t a_{ij}\big)
\right)
+\left(\zeta^{\text{G}}-\frac{2\eta^{\text{G}}}{d}
\right)\left(\theta^{\text{G}}\right)^2
+\kappa^{\text{G}} \Delta T,
\label{galentropy1} 
\end{equation}
where we assumed $\kappa^{\text{G}}$ constant (otherwise the last term would read $\nabla^i(\kappa^{\text{G}} \nabla_i T)$).

A special class of Galilean fluids deserves further analysis. These are the \emph{incompressible fluids} for which $\varrho(t,\mathbf{x})$ obeys
\begin{equation}
\label{galincompr}
\frac{\text{d}\varrho(t,\mathbf{x})}{\text{d}t}=0
\end{equation}
with $\nicefrac{\text{d}}{\text{d}t}$ the material derivative defined in \eqref{galfder}.  Using the  expressions \eqref{galC} and \eqref{galCpart}, we recast the incompressibility requirement as the vanishing of the effective fluid expansion:
 \begin{equation}
\label{galincomprexp}
\theta^{\text{G}}=0.
\end{equation}
In this case, the bulk viscosity drops from the stress tensor \eqref{galsigG1} and the Galilean shear \eqref{galshear} simplifies. The first-order hydrodynamics momentum equation for an incompressible fluid thus reads: 
\begin{equation}
 \varrho \frac{\text{d}\nicefrac{v_i}{\Omega}}{\Omega\,\text{d}t}= \varrho\frac{\text{d}\nicefrac{w_i}{\Omega}}{\Omega\,\text{d}t}+\frac{\varrho}{2}\partial_i \frac{\mathbf{w}^2}{\Omega^2} -\varrho\frac{v^j}{\Omega}\nabla_i\frac{w_j}{\Omega}-\partial_i p +\frac{\eta^{\text{G}}}{\Omega}\left(\Delta v_i +r_ i^{\hphantom{i}j}v_j
+a_{ik} a^{jl}\partial_t\gamma^k_{jl}\right).
\label{NS}
\end{equation} 
We immediately recognize in this expression the generalized \emph{covariant Navier--Stokes equation}, valid for incompressible fluids on any space $\mathscr{S}$, observed from an arbitrary frame. The first three terms in the right-hand side are contributions of frame inertial forces, the fourth is the pressure force, and next come the friction forces at first-order derivative. For Euclidean space with $\Omega=1$ and  $\mathbf{w}=0$ we recover the textbook form
\begin{equation}
\dfrac{\text{d} \mathbf{v}}{\text{d}t}=-\dfrac{\text{\bf grad}\, p}{\varrho}+\dfrac{\eta^{\text{G}}}{\varrho}\Delta \mathbf{v}.
\end{equation}

\subsection{Carrollian fluid dynamics  from Randers--Papapetrou background}\label{FLUID2}

\subsubsection*{Preliminary remarks}

{ As Carrollian particles, Carrollian fluids have no motion. From a relativistic perspective this is an observer-dependent statement, since boosts can turn on velocity. In the limit of vanishing velocity of light, however, these transformations are no longer permitted. Hence, being at rest becomes a genuinely intrinsic feature. 

The fluid velocity must be set to zero faster than  $c$ in order to avoid blow-ups in the energy--momentum conservation. The appropriate scaling, ensuring a non-trivial kinematic contribution is 
\begin{equation}
\label{carbetav}
v^i=c^2 \Omega \beta^i + \text{O}\left(c^4\right),
\end{equation}
where $v^i=\nicefrac{u^i}{\gamma}$.} This 
leaves  the Carrollian fluid with a kinematic variable $\pmb{\beta}=\beta^i\partial_i$ of inverse-velocity dimension, as in \eqref{carbeta} for the one-body Carrollian dynamics studied in Sec. \ref{CARdyn} -- reason why we keep the same symbol.  
In order to reach covariant Carrollian fluid equations by expanding the relativistic fluid equations at small $c$, we need to define the $\beta^i$s in such a way that they transform as components of a genuine Carrollian vector under \eqref{cardifs},  \eqref{carjst} already at finite $c$.
This is achieved by setting 
\begin{equation}
v^{i}=\frac{c^2\Omega\beta^i}{1+c^2\beta^j b_j}
\Leftrightarrow
\beta^{i}=\frac{v^i}{c^2\Omega\left(1-\frac{v^j b_j}{\Omega}\right)},
\end{equation}
from which one checks that\footnote{This is easily proven by observing that
$\beta_i+b_i=-\frac{\Omega u_i}{cu_0}$. We define as usual $b^i=a^{ij}b_j$, $\beta_i = a_{ij}\beta^j$, $v_i = a_{ij}v^j$, 
$\pmb{b}^2=b_ib^i$,  $\pmb{\beta}^2=\beta_i \beta^i$ and $\pmb{b}\cdot \pmb{\beta}=b_i \beta^i$.}
\begin{equation}
\label{car-beta-tran}
\beta^{i\prime}=J^i_j \beta^{j}.
\end{equation}
The full fluid congruence reads then:
\begin{equation}
\label{carvel}
\begin{cases}
\displaystyle{u^0=\gamma c=\dfrac{c}{\Omega}\dfrac{1+c^2 \pmb{\beta}\cdot\pmb{b}}{\sqrt{1-c^2\pmb{\beta}^2}}=\dfrac{c}{\Omega}+\text{O}\left(c^3\right),}
\quad
u_0=-\dfrac{c\Omega}{\sqrt{1-c^2\pmb{\beta}^2}}=-c\Omega+\text{O}\left(c^3\right),
\\
\displaystyle{u^i=\gamma v^i=\dfrac{c^2\beta^i}{\sqrt{1-c^2\pmb{\beta}^2}}=c^2\beta^i+\text{O}\left(c^4\right),
\quad u_i=\dfrac{c^2\left(b_i+\beta_ i\right)}{\sqrt{1-c^2\pmb{\beta}^2}}=c^2\left(b_i+\beta_ i\right)+\text{O}\left(c^4\right) ,}
\end{cases}
\end{equation}
where the Lorentz factor has been obtained by imposing the usual normalization  ${\| \text{u} \|^2=-c^2}$:\begin{equation}
\gamma=\dfrac{1+c^2 \pmb{\beta}\cdot\pmb{b}}{{\Omega}\sqrt{1-c^2\pmb{\beta}^2}}
=\frac{1}{\Omega}\left(1+\frac{c^2}{2}\pmb{\beta}\cdot\left(\pmb{\beta}+2\pmb{b}\right)+\text{O}\left(c^4\right)
\right).
\end{equation}
In the relativistic regime, \emph{i.e.} before taking the zero-$c$ limit, in the Randers--Papapetrou background \eqref{carrp} the perfect part of the energy--momentum tensor
reads then:
\begin{equation}
\label{carTperf} 
\begin{cases}
\displaystyle{T^{\hphantom{\text{perf}}0}_{\text{perf}\hphantom{0}0}=-\varepsilon -c^2(\varepsilon+p) \beta^k \left(b_k+\beta_k\right) + \text{O}\left(c^4\right),}
\\
\displaystyle{c\Omega T^{\hphantom{\text{perf}}0}_{\text{perf}\hphantom{0}i}=c^2(\varepsilon+p) \left(b_i+\beta_i\right)+ \text{O}\left(c^4\right),}
\\
\displaystyle{\dfrac{c }{\Omega}T^{\hphantom{\text{perf}}j}_{\text{perf}\hphantom{j}0}=-c^2(\varepsilon+p)\beta^j + \text{O}\left(c^4\right),}
\\
\displaystyle{T^{\hphantom{\text{perf}}j}_{\text{perf}\hphantom{j}i}=p \delta_{i}^{j} + c^2(\varepsilon+p)\beta^j  \left(b_i+\beta_i\right)+ \text{O}\left(c^4\right).}
\end{cases}
\end{equation}
The non-perfect part is encoded in Eqs. \eqref{T}, \eqref{sigexpC} and \eqref{QexpC}. Notice, on the one hand, that for vanishing $\beta^i$, these expressions are exact at finite $c$: most of the terms of order $c^2$ vanish as do all non-displayed higher-order contributions in $c^2$; on the other hand, for vanishing $c$, one recovers the perfect energy--momentum of a fluid at rest due to the simultaneous vanishing of $v^i$ as a consequence of \eqref{carbetav}.

{ The eventual absence of  motion, macroscopic or microscopic, and the shrinking of the light-cone raise many fundamental questions regarding the origin of pressure, temperature, thermalization, entropy etc. One may wonder in 
particular what causes viscosity and thermal conduction, what replaces the temperature derivative expansion of $q_{i}$, what justifies its behaviour 
\eqref{sigexpC}. Even the propagation of a signal such as sound, if possible,  should be reconsidered.
It is tempting to claim that all this physics will be mostly of geometric nature rather than many-body statistics, because as we will see the only kinematic Carrollian-fluid variable $\pmb{\beta}$ enters partly the dynamics.  

We have no definite answers to all these questions though, and will not discuss these important issues here, which might possibly require to elaborate on space-filling branes as microscopic objects -- see Sec. \ref{CARdyn}. Our approach will be kinematical, aiming at writing the fundamental equations, covariant under Carrollian diffeomorphisms   \eqref{cardifs},  \eqref{carjst}, starting from the relativistic equations \eqref{conT}. Alternative paths may exist, allowing to built some Carrollian dynamics without using the zero-$c$ limit of a relativistic fluid.\footnote{In this spirit, one should quote the attempt made in \cite{Penna3}, inspired by the membrane paradigm --  admittedly suited for reaching Galilean rather than ultra-relativistic fluid dynamics, as well as Ref. \cite{dutch}, mostly  
focused on the structure of the energy--momentum tensor of perfect fluids \eqref{carTperf}, which also touches on Carrollian symmetry.}}

\subsubsection*{The structure of the equations}

The relativistic equations (conservation of the energy--momentum tensor) should now be presented as  
 \begin{equation}
 \label{conTcar} 
\nabla_\mu T^{\mu}_{\hphantom{\mu}0}=0,\quad \nabla_\mu T^{\mu i}=0.
\end{equation}
Under Carrollian diffeomorphisms \eqref{cardifs},  \eqref{carjst}, the divergence of the energy--momentum tensor transforms as:
 \begin{equation}
 \label{conTcarjst} 
\nabla_\mu^{\prime} T^{\prime\mu}_{\hphantom{\prime\mu}0}=\frac{1}{J} \nabla_\mu T^{\mu}_{\hphantom{\mu}0},\quad \nabla_\mu^{\prime} T^{\prime\mu i}=J_{l}^{i}  \nabla_\mu T^{\mu l}.
\end{equation}
In analogy with the Galilean case  \eqref{conTgal}, the two sets of equations \eqref{conTcar} have separately a $d$-dimensional covariant transformation. This is part of the agenda for the Carrollian dynamics.

Equations \eqref{conTcar} are relativistic. Using the general energy--momentum tensor \eqref{T} with perfect part \eqref{carTperf} and \eqref{sigexpC} as stress tensor, we find generally:
 \begin{eqnarray}
 \label{conTcarexp0} 
\frac{c}{\Omega}\nabla_\mu T^{\mu}_{\hphantom{\mu}0}&=& \frac{1}{c^2}{\mathcal{F}}
+ {\mathcal{E}}
+\text{O}\left(c^2\right),
\\
\label{conTcarexpi} 
 \nabla_\mu T^{\mu i}&=&\frac{1}{c^2} \mathcal{H}^i
 +\mathcal{G}^i
 +\text{O}\left(c^2\right) .
\end{eqnarray}
For zero $\beta^i$, these expressions are \emph{exact}
with extra terms of order $c^2$ only, and requiring they vanish leads to the  $d+1$
fully relativistic fluid equations. With $\beta^i\neq 0$, \eqref{conTcarexp0} and  \eqref{conTcarexpi} 
are genuinely infinite series. Thanks to the validity of \eqref{car-beta-tran} at finite $c$, Carrollian diffeomorphisms do not mix the different orders of these series, making each term Carrollian-covariant.
Here, we are interested in the zero-$c$ limit, and in this case Eqs. \eqref{conTcarexp0} and  \eqref{conTcarexpi}  split into $2+2d$ distinct equations:
\begin{itemize}
\item energy conservation $\mathcal{E}=0$;
\item momentum conservation $\mathcal{G}^i=0$;
\item constraint equations $\mathcal{F}=0$ and $\mathcal{H}^i=0$.
\end{itemize}
All of these are covariant under Carrollian diffeomorphisms  \eqref{cardifs},  \eqref{carjst}.  
 
The Carrollian fluid, obtained as Carrollian limit of a relativistic fluid in the appropriate  
(Randers--Papapetrou) background, is described in terms of the $d$  $\beta^i$s, 
and the
two variables $p$ and $\varepsilon$.\footnote{The proper energy density cannot be split in mass density and energy per mass, because  the limit at hand is ultra-relativistic. Observe also that $\pmb{b}$ is not a fluid variable but a Carrollian-frame parameter as was $\mathbf{w}$ in the Galilean case. The fluid kinematical variable is $\pmb{\beta}$, playing the r\^ole $\frac{\mathbf{v}-\mathbf{w}}{\Omega}$ had in the usual non-relativistic case.} The latter are related through an equation of state and the energy-conservation equation $\mathcal{E}=0$. As we will see soon, the other 
$2d + 1$ equations are setting consistency constraints among the $2d$ components  
of the heat currents ($Q^{\text{C}}_{\hphantom{\text{C}}i}$ and $\pi_i$), the $d(d+1)$ components of the viscous stress tensors ($\Sigma^{\text{C}}_{\hphantom{\text{C}}ij}$ and $\Xi_{ij}$), the inverse-velocity components $\beta^i$ and the geometric environment. Geometry is therefore expected to interfere more actively in the dynamics of Carrollian fluids,
than it did for Galilean hydrodynamics. Some of the aforementioned constraints are possibly rooted to more fundamental microscopic/geometric properties, yet to be unravelled. Their usage will be illustrated in Sec. \ref{EXART}.

\subsubsection*{The dissipative tensors in Randers--Papapetrou background}

For a relativistic fluid in the Randers--Papapetrou background \eqref{carrp}, using the velocity field in \eqref{carbetav} and  \eqref{carvel} and the components $q^i$, the transversality conditions \eqref{trans} lead to 
\begin{equation}
 \label{carheat}
q^0=\frac{c}{\Omega}\left(b_i+ \beta_i\right) q^i
 ,\quad
 q_0=-c\Omega \beta_i q^i
 ,\quad
q_i=\left(a_{ij}
+c^2b_i  \beta_j \right)q^j.
\end{equation}
Similarly, the components of the viscous stress tensor are obtained from the $\tau^{ij}$s. For example:
\begin{equation}
 \label{carfs} 
\begin{array}{rcl}
&\displaystyle{\tau^{00}= \frac{c^2}{\Omega^2}\left(b_k+ \beta_k\right)\left(b_l+ \beta_l\right) \tau^{kl},
\quad
\tau^{0i}= \frac{c}{\Omega}\left(b_i+ \beta_i\right) \tau^{ik},
\quad
 \tau_{00}=c^2\Omega^2\beta_k\beta_l\tau^{kl},}
\\
&\displaystyle{\tau_{0i}=- c\Omega\beta_j\left(
a_{ik}
+c^2b_{i\vphantom{j}}  \beta_{k} 
\right)\tau^{jk},
\quad 
\tau_{ij}=
\left(
a_{ik}
+c^2b_{i\vphantom{j}}  \beta_{k} 
\right)\left(
a_{jl}
+c^2b_j  \beta_l 
\right)\tau^{kl},
\ldots}
\end{array}
\end{equation}

Under Carrollian diffeomorphisms  \eqref{cardifs},  \eqref{carjst}, we obtain the following transformation rules
\begin{equation}
\label{cardifqtau}
q^{\prime i} =q^{j} J^{i}_{j}, \quad \tau^{\prime ij} =\tau^{kl} J_{k}^{i} J_{l}^{j} .
\end{equation}
This suggests to use $q^i$  as components for the Carrolian $d$-dimensional heat current decomposed as 
$Q^{\text{C} i}+c^2 \pi^i$ (see \eqref{QexpC}),
and 
$\tau^{ij} $  for the Carrolian $d$-dimensional viscous stress tensors
$\Sigma^{\text{C}ij}$ and  $\Xi^{ij}$ defined in \eqref{sigexpC}. We introduce as usual
\begin{eqnarray}
 \label{carheatdsf} 
&Q^{\text{C}}_{\hphantom{\text{C}}i}
= a_{ij}
Q^{\text{C}j}
,
\quad
\Sigma^{\text{C}\hphantom{i}j}_{\hphantom{\text{C}}i}
= a_{ik}
\Sigma^{\text{C}kj} 
,\quad
\Sigma^{\text{C}}_{\hphantom{\text{C}}ij}
= 
a_{jk}
\Sigma^{\text{C}\hphantom{i}k}_{\hphantom{\text{C}}i}
,&
\\
 \label{carheatdsfppsi} 
&\pi_{i}= a_{ij}\pi^{j},\quad
\Xi^{\hphantom{i}j}_{i}
= a_{ik}
\Xi^{kj} 
,\quad
 \Xi_{ij}=a_{jk}
\Xi^{\hphantom{i}k}_{i}.&
\end{eqnarray}

Using the generic transformations \eqref{cardifqtau} 
under Carrollian diffeomorphisms \eqref{cardifs},  \eqref{carj}, we find that the above quantities transform as they should, for being eligible as $d$-dimensional tensors:
\begin{eqnarray}
 \label{carheatddiff} 
&Q^{\text{C}\prime }_{\hphantom{\prime}i} = Q^{\text{C}}_{\hphantom{\text{C}}{j}} J^{-1j}_{\hphantom{-1}i}, 
\quad
Q^{\text{C}\prime i}=J^{i}_{j}Q^{\text{C}j},&
\\
 \label{carfsddiff} 
& \Sigma^{\text{C}\prime }_{\hphantom{\prime \text{C}}ij} 
=J^{-1k}_{\hphantom{-1}i}J^{-1l}_{\hphantom{-1}j}\Sigma^{\text{C}}_{\hphantom{\text{C}}kl}, 
\quad
\Sigma^{\text{C}\prime\hphantom{i}j}_{\hphantom{\prime\text{C}}i}
=J^{-1k}_{\hphantom{-1}i} \Sigma^{\text{C}\hphantom{k}l}_{\hphantom{\text{C}}k} J^{j}_{l}, \quad
\Sigma^{\text{C}\prime ij} =\Sigma^{\text{C} kl} J^{i}_{k}J^{j}_{l},&
\end{eqnarray}
and similarly for $\pi_{i}$ and $ \Xi_{jk}$.

\subsubsection*{Scalar equations}

The computation of the spacetime divergence in \eqref{conTcarexp0} is straightforward and leads to the following:
\begin{eqnarray}
\mathcal{E}&=&
- \left(\frac{1}{\Omega}\partial_t +\frac{d+1}{d}\theta^{\text{C}}\right)\left(\varepsilon
+2\beta_iQ^{\text{C}i}
-\beta_i\beta_j\Sigma^{\text{C}ij}
\right)+\frac{1}{d}\theta^{\text{C}}
\left(\Xi^{i}_{\hphantom{i}i}-\beta_i \beta_j \Sigma^{\text{C}ij}+\varepsilon - d p\right)
\nonumber
\\
&&-\left(\hat\nabla_i +2\varphi_i \right)\left(Q^{\text{C}i}
-\beta_j \Sigma^{\text{C}ij}\right)
-\left(2Q^{\text{C}i}\beta^j-\Xi^{ij} \right)\xi^{\text{C}}_{\hphantom{\text{C}}ij},
 \label{carE} 
\\
 \label{carF} 
\mathcal{F}&=&\Sigma^{\text{C}ij}\xi^{\text{C}}_{ij}+\frac{1}{d}\Sigma^{\text{C}i}_{\hphantom{\text{C}i}i}\theta^{\text{C}},
\end{eqnarray}
where we have introduced a new covariant derivative $\hat\nabla_i$, as defined in the appendix, Eqs. \eqref{dgammaCar}--\eqref{Carcovde2ten}. It is based on a new torsionless and metric-compatible connection (see \eqref{carconmet}--\eqref{carcontor}) dubbed \emph{Levi--Civita--Carroll}, which plays for Carrollian geometry the r\^ole of  
ordinary Levi--Civita connection for ordinary geometry,  \emph{i.e.} it allows to built derivatives covariant under Carrollian diffeomorphisms  \eqref{cardifs},  \eqref{carj}. Some further properties regarding the curvature of this connection are displayed in \eqref{carriemann}--\eqref{carriemanntimetilde}.  A deeper investigation of this structure is  out of place here. In \eqref{carE} and \eqref{carF} we have moreover defined
\begin{eqnarray}
\label{caracc} 
\varphi_i&=&\dfrac{1}{\Omega}\left(\partial_t b_i+\partial_i \Omega\right),
\\
\label{carexp1} 
\theta^{\text{C}}&=&
\dfrac{1}{\Omega}              
\partial_t \ln\sqrt{a}.
\end{eqnarray}
These expressions describe a form and a scalar (see \eqref{carb1} and \eqref{cara2} for their transformation rules under Carrollian diffeomorphisms). They play the r\^ole of \emph{inertial acceleration} and \emph{expansion} for the Carrollian fluid. These are both geometrical and the qualifier ``inertial'' refers to the frame (\emph{i.e.} $b_i$ and $\Omega$) origin. We shall see in a moment that there is an extra contribution to the Carrollian fluid acceleration due to the kinematical observable $\beta_ i$, but none for the expansion  (see \eqref{caracclim}, \eqref{carexp}). As already stated and readily seen by its equations, most of the fluid properties are of geometrical nature. One similarly defines an \emph{ inertial vorticity two-form} with components
\begin{equation}
\label{carom}
\varpi_{ij}=\partial_{[i}b_{j]}+b_{[i}\varphi_{j]},
\end{equation}
and the traceless and symmetric \emph{shear tensor}
\begin{equation}
\label{carsh}
\xi^{\text{C}}_{\hphantom{\text{C}}ij}=\dfrac{1}{\Omega}\left(\dfrac{1}{2} \partial_t a_{ij}-
\dfrac{1}{d} a_{ij} \partial_t \ln\sqrt{a}\right).
\end{equation}
These quantities will be related in a short while to the ordinary relativistic counterparts (see \eqref{caromlim} and \eqref{carsiglim}). The former receives a fluid kinematical contribution, as opposed to the latter. Eventually, we can elegantly check that
\begin{equation}
\label{carEFdif} 
\mathcal{E}^\prime=  \mathcal{E},\quad \mathcal{F}^\prime=\mathcal{F}
\end{equation}
(we use for that Eqs. \eqref{cardifa}, \eqref{carheatddiff}, 
\eqref{carfsddiff}, \eqref{carb1},  \eqref{carb2},  \eqref{Carcovdersca}--\eqref{car2}). 

Equation $\mathcal{F}=0$ sets a geometrical constraint on the Carrollian stress tensor $\pmb{\Sigma}^\text{C}$, whereas $\mathcal{E}=0$ is the energy conservation. Using  \eqref{carE}, the latter can be recast as follows:
 \begin{equation}
 \left(\frac{1}{\Omega}\partial_t +\theta^{\text{C}}\right) e_{\text{e}}=
 -\left(\hat\nabla_i +2\varphi_i \right)\Pi^{\text{C}i}-\Pi^{\text{C}ij}\left(  \xi^{\text{C}}_{\hphantom{\text{C}}ij} +\frac{1}{d}\theta^{\text{C}}a_{ij} \right),
  \label{carEbis} 
\end{equation}
and in this form it bares some resemblance with the Galilean homologous equation \eqref{galEEuler}. 
It exhibits three Carrollian tensors, which capture the Carrollian energy exchanges:
\begin{equation}
e_{\text{e}}=
\varepsilon
+2\beta_iQ^{\text{C}i}
-\beta_i\beta_j\Sigma^{\text{C}ij}, \quad 
\Pi^{\text{C}i}
=Q^{\text{C}i}
-\beta_j \Sigma^{\text{C}ij},\quad
\Pi^{\text{C}ij}
=Q^{\text{C}i}\beta^j+\beta^iQ^{\text{C}j}+pa^{ij}-\Xi^{ij}.
\label{carflux}
\end{equation}
The first is a scalar $e_{\text{e}}$, which can be interpreted as an \emph{effective Carrollian energy density} (observe the absence of kinetic energy, expected from the vanishing velocity).
Its time variation, including the dilution/contraction effects due to the expansion, is driven  by the gradient of a \emph{Carrollian energy flux}, which is the vector $\Pi^{\text{C}i}$, and by the coupling of the shear to a \emph{Carrollian energy--momentum tensor} $\Pi^{\text{C}ij}$.

\subsubsection*{Vector equations}

The vectorial part of the divergence is obtained from \eqref{conTcarexpi} and has two pieces.
The first reads:
\begin{eqnarray}
\mathcal{G}_j&=&
\left(\hat\nabla_i+\varphi_i\right)\Pi^{\text{C}i}_{\hphantom{\text{C}i}j} +\varphi_j e_{\text{e}} +2
\Pi^{\text{C}i}\varpi_{ij}
+\left(\dfrac{1}{\Omega}\partial_t+\theta^{\text{C}}\right)\left(\pi_j+\beta_j\left(
e_{\text{e}}
-2\beta_i\Pi^{\text{C}i}
-\beta_i\beta_k\Sigma^{\text{C}ik}
\right)
\right)
\nonumber\\
&& +\left(\dfrac{1}{\Omega}\partial_t+\theta^{\text{C}}\right)\left(\beta^k\left(\Pi^{\text{C}}_{\hphantom{\text{C}}kj}
-\dfrac{1}{2}\beta_k\Pi^{\text{C}}_{\hphantom{\text{C}}j}-\dfrac{1}{2}\beta_k \beta^i
\Sigma^{\text{C}}_{\hphantom{\text{C}}ij}
  \right)\right).
  \label{carG} 
  \end{eqnarray}
The second is as follows:
 \begin{equation}
  \mathcal{H}_j=
- \left(\hat\nabla_i + \varphi_i\right)\Sigma^{\text{C}i}_{\hphantom{\text{C}i}j}+
\left(\frac{1}{\Omega}\partial_t +\theta^{\text{C}}
\right)\Pi^{\text{C}}_{\hphantom{\text{C}}j}.
 \label{carH} 
\end{equation}
Equation $\mathcal{G}_j=0$ involves $\varepsilon$, $p$ and
their temporal and/or spatial derivatives, $\pmb{\beta}$, the heat current $\pmb{Q}^\text{C}$,  and $\pmb{\Xi}$, expressed in terms of the effective energy density $e_{\text{e}}$, the Carrollian energy flux and energy--momentum tensor $\pmb{\Pi}^\text{C}$, as well as $\pmb{\pi}$ and  $\pmb{\Sigma}^\text{C}$.  It is a momentum conservation. Notice also the coupling of the energy flux to the inertial vorticity.
Equation $\mathcal{H}_j=0$ depends neither on $\varepsilon$ nor on $p$. This is an equation for the Carrollian energy flux $\pmb{\Pi}^\text{C}$ and the viscous stress tensor $\pmb{\Sigma}^\text{C}$, of geometrical nature as it involves the metric $\pmb{a}$, the Carrollian ``frame velocity''~$\pmb{b}$ and the inertial acceleration $\pmb{\varphi}$.

Under Carrollian diffeomorphisms   \eqref{cardifs},  \eqref{carj}, using the already quoted equations,  \eqref{cardifa}, \eqref{carheatddiff}, 
\eqref{carfsddiff}, and \eqref{carb1}--\eqref{car3}, we obtain:
\begin{equation}
\label{carvdif} 
\mathcal{G}^{\prime i} = J^{i}_{j} \mathcal{G}^j,\quad\mathcal{H}^{\prime i} = J^{i}_{j} \mathcal{H}^j.
\end{equation}

One should observe at this point that the energy--momentum tensor and energy flux associated with a Carrollian fluid and defined in \eqref{carflux} are merely a repackaging of part of the dynamical data. They do not capture all perfect and friction quantities, as it happens for Galilean fluids, Eqs. \eqref{galPI} and \eqref{galMEnflux}. 
Equation $\mathcal{F}=0$, as well as the vector equations need indeed more information than the energy--momentum tensor and energy flux.
There is pressure, energy density and ``velocity'', on the one hand, and on the other hand, we find the two heat currents and the two viscous stress tensors. The zero-$c$ limit produces a decoupling in the equations, sustained by the scaling assumption \eqref{sigexpC}. This is the reason why $\mathcal{H}_j=0$ appears as an equation for the dissipative pieces of data only, while the non-dissipative ones mix with the heat currents inside $\mathcal{G}_j=0$.

\subsubsection*{Carrollian perfect fluids}

We would like to end this chapter with a remark on the case of perfect fluids, namely fluids with vanishing dissipative tensors. For those, the dynamical variables are $\varepsilon$, $p$ and $\beta_i$, with $e_{\text{e}}=\varepsilon$, $\Pi^{\text{C}}_{\hphantom{\text{C}}j}=0$ and 
$\Pi^{\text{C}}_{\hphantom{\text{C}}ij}=p a_{ij}$. In this case, $\mathcal{F}=\mathcal{H}^i=0$ identically, and 
\begin{eqnarray}
\mathcal{E}&=&
- \frac{1}{\Omega}\partial_t \varepsilon-(\varepsilon+p)\theta^{\text{C}}\,,
 \label{carEp} 
\\
\mathcal{G}_j&=&
(\varepsilon+p)\left(\varphi_j+\gamma^{\text{C}}_{\hphantom{C}j}+\beta_j \theta^{\text{C}}\right)
+\frac{\beta_j}{\Omega}\partial_t(\varepsilon+p)
+\hat\partial_j p.
 \label{carGp} 
\end{eqnarray}
On the one hand, non-trivial energy exchanges can only result from time-dependence of the metric and pressure gradients. The latter, on the other hand, are bound to non-trivial $\pmb{\beta}$, $\pmb{\gamma}^{\text{C}}$, $\pmb{b}$ and $\Omega$. Here $\gamma^{\text{C}}_{\hphantom{C}j}$ is the kinematical acceleration defined later in \eqref{carkinacc}.

For perfect fluids, only $\mathcal{E}$ and $\mathcal{G}_i$ survive in the relativistic divergence of the energy--momentum tensor, Eqs. \eqref{conTcarexp0} and \eqref{conTcarexpi}. Furthermore, for zero $\pmb{\beta}$ these are actually the only terms, at finite $c$. Hence,
the relativistic equations are not affected by the vanishing-$c$ limit, and coincide with the Carrollian ones:
$\mathcal{E}=0$ and $\mathcal{G}_i=0$. As a consequence, the Carrollian nature of a fluid at $\pmb{\beta}=0$ can only emerge through interactions.  This is to be opposed to the Galilean situation, since Galilean perfect fluids are definitely different from relativistic perfect fluids, even at rest. 

\subsubsection*{First-order Carrollian hydrodynamics}

In order to acquire a better perspective of Carrollian fluid dynamics, we can study the first-order in derivative expansion of its viscous tensors and heat currents. The first-derivative relativistic kinematical tensors as 
acceleration and expansion \eqref{def21},
shear  \eqref{def23},  and vorticity \eqref{def24}, for a fluid with velocity vanishing as \eqref{carbetav} when $c\to 0$ in Randers--Papapetrou background \eqref{carrp} read (the only independent components are the spatial ones):
\begin{eqnarray}
\label{caracclim} 
&&a_i=\dfrac{c^2}{\Omega}\left(\partial_t \left(b_i+\beta_i\right)+\partial_i \Omega\right)+\text{O}\left(c^4\right)=c^2\left(\varphi_i+\gamma^{\text{C}}_{\hphantom{C}i}\right) +\text{O}\left(c^4\right),
\\
\label{carexp} 
&&\Theta=
\dfrac{1}{\Omega}
\partial_t \ln\sqrt{a}+\text{O}\left(c^2\right)
=\theta^{\text{C}}+\text{O}\left(c^2\right),
\\
\label{carsiglim} 
&&\sigma_{ij}=\dfrac{1}{\Omega}\left(\dfrac{1}{2} \partial_t a_{ij}-
\dfrac{1}{d} a_{ij} \partial_t \ln\sqrt{a}\right)+\text{O}\left(c^2\right)
=
\xi^{\text{C}}_{\hphantom{\text{C}}ij}+\text{O}\left(c^2\right),
\\
\label{caromlim} 
&&\omega_{ij}=c^2\left(\partial_{[i}b_{j]}+\frac{1}{\Omega}b_{[i}\partial_{j]}\Omega+\frac{1}{\Omega}b_{[i}\partial_tb_{j]}+w_{ij}\right)+\text{O}\left(c^4\right)=c^2\left(\varpi_{ij}+w_{ij}\right)+\text{O}\left(c^4\right)
.\qquad
\end{eqnarray}
We find the corresponding Carrollian expansion  $\theta^{\text{C}}$ and  shear $\xi^{\text{C}}_{\hphantom{\text{C}}ij}$, as already anticipated in \eqref{carexp1} and \eqref{carsh}. These quantities are purely geometric and originate from the time dependence of the $d$-dimensional spatial metric. Similarly, the relativistic acceleration and vorticity allow to define the already introduced Carrollian, inertial acceleration $\varphi_i$ and vorticity $\varpi_{ij}$, as well as the kinematical acceleration $\gamma^{\text{C}}_{\hphantom{C}i}$ and kinematical vorticity $w_{ij}$ defined as:
\begin{eqnarray}
\label{carkinacc} 
\gamma^{\text{C}}_{\hphantom{C}i}&=&\dfrac{1}{\Omega}\partial_t \beta_i,
\\
\label{carkinvort} 
w_{ij}&=&\hat\partial_{[i}\beta_{j]}+\beta_{[i}\varphi_{j]}+ 
\beta^{\vphantom{[}}_{[i}\gamma^{\text{C}}_{\hphantom{C}{j]}}.
\end{eqnarray}
Starting from the first-order relativistic viscous tensor \eqref{e1} and heat current  \eqref{q1}, 
in order to comply with the behaviours \eqref{sigexpC} and the definition of the Carrollian heat currents \eqref{QexpC}, we must assume that (up to possible higher orders in $c^2$)
\begin{equation}
\label{relcartr}
\eta=\tilde \eta+\frac{\eta^{\text{C}}}{c^2},\quad
\zeta=\tilde \zeta+\frac{\zeta^{\text{C}}}{c^2},\quad 
\kappa=c^2 \tilde \kappa++\kappa^{\text{C}}.
\end{equation}
Hence, putting these equations together, we find 
\begin{eqnarray}
\label{Car1stordertau2}
\Sigma_{(1)ij}^{C}&=& 2\eta^{\text{C}}\xi_{\hphantom{\text{C}}ij}^{C}+\zeta^{\text{C}}\theta^C a_{ij},
\\
\nonumber
Q^{\text{C}}_{(1)i}&=& -\frac{\kappa^{\text{C}}}{\Omega}\left(\partial_t(b_i T)+\beta_i \partial_t T+\partial_i(\Omega T)\right)
\\&=&
-\kappa^{\text{C}}\left(
\hat\partial_i T +T\left(\varphi_i+\gamma^{\text{C}}_{\hphantom{C}i}\right)
\right),
\label{Car1storderq}
\end{eqnarray}
and similarly for $\Xi_{(1)ij}$ and $\pi_{(1)i}$. These quantities will include respectively terms like 
$2\tilde\eta\xi_{\hphantom{\text{C}}ij}^{C}+\tilde\zeta\theta^C a_{ij}  $
and $-\tilde\kappa\left(
\hat\partial_i T +T\left(\varphi_i+\gamma^{\text{C}}_{\hphantom{C}i}\right)
\right)$, plus extra terms coupled to  $\eta^{\text{C}}$, $\zeta^{\text{C}}$ and $\kappa^{\text{C}}$, and originating from higher-order contributions in the $c^2$-expansion of the relativistic shear, acceleration and expansion. Notice that these are absent for vanishing $\beta^i$ because in this case \eqref{caracclim}--\eqref{caromlim} are exact. 

All the above expressions are covariant under Carrollian diffeomorphisms \eqref{cardifs}, \eqref{carj} (see formulas \eqref{cara1}--\eqref{carb2}  in appendix).
The friction phenomena are geometric and due to time evolution of the background metric $a_{ij}$. The
heat conduction, depends also on a temperature, which has not been defined in Carrollian thermodynamics due to the absence of kinetic theory. 

In the two-dimensional case  one should take into account the Hall viscosity \eqref{relHj} in the relativistic viscous tensor at first order. Assuming again $\zeta_{\text{H}}=\nicefrac{\zeta_{\text{H}}^{\text{C}}}{c^2}+\tilde\zeta_{\text{H}}$, the extra term to be added to $\Sigma_{(1)ij}^{C}$ in \eqref{Car1stordertau2} reads:
\begin{equation}
\label{Car1storderHall}
\zeta_{\text{H}}^{\text{C}}\sqrt{a}\epsilon^{\vphantom{\text{C}}}_{k(i}\xi^C_{\hphantom{\text{C}}j)l}a^{kl},
\end{equation}
and similarly for 
$\Xi_{(1)ij}$ with transport coefficients $\tilde\zeta_{\text{H}}$ and $\zeta_{\text{H}}^{\text{C}}$ as already explained.

The final first-order Carrollian equations are obtained by substituting  $\Sigma_{(1)ij}^{C}$ and
$Q^{\text{C}}_{(1)i}$  
given in \eqref{Car1stordertau2}
and   \eqref{Car1storderq},
and similarly for 
$\Xi_{(1)ij}$, 
and
$\pi_{(1)i}$, inside the general expressions for $\mathcal{E}$, 
$\mathcal{F}$, 
$\mathcal{G}_i$ and
$\mathcal{H}_i$, Eqs. \eqref{carE},  \eqref{carF},  \eqref{carG} and  \eqref{carH}.

\subsubsection*{Conformal Carrollian fluids}

Carrollian fluids are ultra-relativistic and are thus compatible with conformal symmetry.
For conformal
relativistic fluids the energy--momentum tensor \eqref{T} is traceless and this requires
\begin{equation}\label{con}
\varepsilon= dp,\quad \tau^\mu_{\hphantom{\mu}\mu}=0. 
\end{equation} 
In the Carrollian limit, the latter reads:
\begin{equation}\label{con-carroll}
\Xi^{i}_{\hphantom{i}i}=\beta_i \beta_j \Sigma^{\text{C}ij},\quad \Sigma^{\text{C}i}_{\hphantom{\text{C}i}i}=0. 
\end{equation} 
In particular, we find $e_{\text{e}}=\Pi^{\text{C}i}_{\hphantom{\text{C}i}i}$.

The dynamics of conformal fluids is covariant under Weyl transformations. Those act on the fluid variables as
\begin{equation}
\label{weyl-fluid}
\varepsilon \to \mathcal{B}^{d+1}\varepsilon,\quad \pi_i\to \mathcal{B}^{d} \pi_i,\quad
 Q^{\text{C}}_{\hphantom{\text{C}}i}\to \mathcal{B}^{d}Q^{\text{C}}_{\hphantom{\text{C}}i},\quad \Xi_{ij}\to \mathcal{B}^{d-1} \Xi_{ij},\quad
 \Sigma^{\text{C}}_{\hphantom{\text{C}}ij}\to \mathcal{B}^{d-1} \Sigma^{\text{C}}_{\hphantom{\text{C}}ij},
\end{equation} 
where $\mathcal{B}=\mathcal{B}(t,\mathbf{x})$ is an arbitrary function. The elements of the Carrollian geometry behave as follows:
\begin{equation}
\label{weyl-geometry}
a_{ij}\to \frac{1}{\mathcal{B}^2}a_{ij},\quad b_{i}\to \frac{1}{\mathcal{B}}b_{i},\quad \Omega\to \frac{1}{\mathcal{B}}\Omega,
\end{equation} 
and similarly for the kinematical variable $\beta_i$, the inertial and kinematical vorticity \eqref{carom} and the shear \eqref{carsh}:
 \begin{equation}
 \label{weyl-geometry-2}
 \beta_i\to \frac{1}{\mathcal{B}} \beta_i, \quad 
\varpi_{ij}\to \frac{1}{\mathcal{B}}\varpi_{ij},\quad 
 w_{ij}\to \frac{1}{\mathcal{B}}w_{ij},\quad
\xi^\text{C}_{\hphantom{\text{C}}ij}\to \frac{1}{\mathcal{B}}\xi^\text{C}_{\hphantom{\text{C}}ij}.
\end{equation}
The Carrollian inertial and kinematical accelerations \eqref{caracc} and \eqref{carkinacc}, and the Carrollian expansion \eqref{carexp1} transform as connections:
 \begin{equation}
 \label{weyl-connection}
\varphi_{i}\to \varphi_{i}-\hat\partial_i\ln \mathcal{B}, \quad \gamma^{\text{C}}_{\hphantom{C}{i}}\to \gamma^{\text{C}}_{\hphantom{C}{i}}-\frac{\beta_i}{\Omega}\partial_t \ln \mathcal{B},\quad \theta^\text{C}\to \mathcal{B}\theta^\text{C}-\frac{d}{\Omega}\partial_t \mathcal{B}.
\end{equation} 
The first and the latter enable to define Weyl--Carroll covariant derivatives $\hat{\mathscr{D}}_i$ and 
$\hat{\mathscr{D}}_t$, as discussed in App. \ref{rpapp}, Eqs. \eqref{CWs-Phi}--\eqref{CWt-met}. With these derivatives, Carrollian expressions  \eqref{carE},  \eqref{carF},  \eqref{carG} and  \eqref{carH} read for a conformal fluid:
\begin{eqnarray}
\mathcal{E}&=&-\frac{1}{\Omega}\hat{\mathscr{D}}_te_{\text{e}}
-\hat{\mathscr{D}}_i \Pi^{\text{C}i}
-\Pi^{\text{C}ij}\xi^{\text{C}}_{\hphantom{\text{C}}ij},
 \label{carEcon} 
\\
\label{carFcon} 
\mathcal{F}&=&
\Sigma^{\text{C}ij}\xi^{\text{C}}_{\hphantom{\text{C}}ij},
\\
\mathcal{G}_j&=& \hat{\mathscr{D}}_i \Pi^{\text{C}i}_{\hphantom{\text{C}i}j}+2\Pi^{\text{C}i}\varpi_{ij}+ \left(\frac{1}{\Omega}\hat{\mathscr{D}}_t \delta^i_j +\xi^{\text{C}i}_{\hphantom{\text{C}i}j}\right)\left(\pi_i+\beta_i\left(
e_{\text{e}}
-2\beta_k\Pi^{\text{C}k}
-\beta_k\beta_l\Sigma^{\text{C}kl}
\right)\right)\nonumber\\
&&+ \left(\frac{1}{\Omega}\hat{\mathscr{D}}_t \delta^i_j +\xi^{\text{C}i}_{\hphantom{\text{C}i}j}\right)\left(\beta^k\left(\Pi^{\text{C}}_{\hphantom{\text{C}}ki}
-\dfrac{1}{2}\beta_k\Pi^{\text{C}}_{\hphantom{\text{C}}i}-\dfrac{1}{2}\beta_k \beta^l
\Sigma^{\text{C}}_{\hphantom{\text{C}}li}
  \right)\right),
  \label{carGcon}\\
\mathcal{H}_j&=&- \hat{\mathscr{D}}_i \Sigma^{\text{C}i}_{\hphantom{\text{C}i}j}
+ \frac{1}{\Omega}\hat{\mathscr{D}}_t 
\Pi^{\text{C}}_{\hphantom{\text{C}}j}
+ \Pi^{\text{C}}_{\hphantom{\text{C}}i}\xi^{\text{C}i}_{\hphantom{\text{C}i}j}.
 \label{carHcon} 
\end{eqnarray}
These equations are Weyl-covariant of weights $d+2$, $d+2$, $d+1$ and $d+1$.

The case of conformal Carrollian perfect fluids is remarkably simple. As quoted earlier $\mathcal{F}=\mathcal{H}^i=0$, and here
\begin{equation}
\label{confo-perf-car}
\mathcal{E}=
- \frac{1}{\Omega}\hat{\mathscr{D}}_t\varepsilon
,\quad
\mathcal{G}_j=\frac{1}{d}\hat{\mathscr{D}}_j \varepsilon
+\frac{d+1}{d}\left(\frac{1}{\Omega}\hat{\mathscr{D}}_t \delta^i_j +\xi^{\text{C}i}_{\hphantom{\text{C}i}j}\right) \varepsilon \beta_i.
\end{equation} 
For these fluids the energy density is covariantly constant with respect to the Weyl--Carroll time derivative.

\subsection{A self-dual fluid}

A duality relationship between the Zermelo and the Randers--Papapetrou background metrics exist and can be stated as follows \cite{Gibbons:2008zi}: the contravariant form of Zermelo matches the covariant expression of Randers--Papapetrou and vice-versa (see Eqs. \eqref{compzerm} and \eqref{comprp}). 

This property is actually closely related to the duality among the Galilean and Carrollian contractions of the Poincar\'e group \cite{Duval:2014uoa}, and has many simple manifestations. For example, the reduction of a spacetime vector representation with respect to Galilean diffeomorphisms \eqref{galdifs}, \eqref{galj}, \eqref{galjst} is performed with the components $V^0$ and $V_i$. Indeed, these transform as
\begin{equation}
V^{\prime 0}=J V^0,\quad V^{\prime}_{i} =V_k J^{-1k}_{\hphantom{-1}i} .
\end{equation}
When reducing under Carrollian diffeomorphisms \eqref{cardifs}, \eqref{carj}, \eqref{carjst}, one should instead use $V_0$ and $V^i$ since 
\begin{equation}
V^{\prime}_{0}=\frac{1}{J} V_0,\quad V^{\prime i} =J_{k}^{i}V^k  .
\end{equation}

The remarkable values $w^i=b_i=0$ and $\Omega=1$ define a sort of self-dual background. If furthermore we require the fluid to be at rest, no distinction survives between \emph{perfect} Galilean and Carrollian fluids, as one readily checks that their equations are identical. The velocity of light is immaterial in this case. As soon as the system is driven away from perfection, this property does not hold any longer, because interactions are sensitive to $c$.  

\section{Examples} \label{EXA}

We will now illustrate our general formalism with examples for Galilean and Carrollian fluids. The latter is the first instance of a fluid obeying exact Carrollian dynamics. It is important both mathematically, as it makes contact with Calabi flows, and physically, for it is relevant in gravity and holography.

\subsection{Galilean fluids}  \label{EXArot}

We provide here two applications: the flat space in rotating frame, which is well known and has the virtue of giving confidence to our methods, and the inflating space, combining both time-dependence and non-flatness of the host $\mathscr{S}$. 

\subsubsection*{Euclidean three-dimensional space in rotating frame}  

We will present the hydrodynamical equations for a non-perfect fluid  moving in Euclidean space $E_3$ with Cartesian coordinates, and observed from a uniformly rotating frame (see \eqref{E3-rot}):
\begin{equation}
\label{E3-unirot}
a_{ij}=\delta_{ij},\quad \Omega = 1,\quad 
\mathbf{w}(\mathbf{x})= \mathbf{x}\times\pmb{\omega}.
\end{equation}
For this fluid, the continuity equation is simply
\begin{equation}
\dfrac{\text{d}\varrho}{\text{d}t}+\varrho\,  \textbf{div}\, \mathbf{v}=0.
\end{equation}

The Euler equation in first-order hydrodynamics, Eq. \eqref{galMEuler1} reads:
\begin{equation}
\label{galMEuler1-rot} 
\dfrac{\text{d}\mathbf{v}}{\text{d}t}=
(\pmb{\omega}\times  \mathbf{x})\times \pmb{\omega}+2\mathbf{v}\times\pmb{\omega}
-\frac{\text{\bf grad}\, p}{\varrho} +\frac{\eta^{\text{G}}}{\varrho}\Delta \mathbf{v} +\frac{1}{\varrho}\left(\zeta^{\text{G}}+\frac{\eta^{\text{G}}}{3}
\right)\text{\bf grad}(\text{\bf div}\,\mathbf{v}),
\end{equation}
and we recognize the various, already spelled  contributions to the dynamics. This equation has been obtained and used in many instances, see \emph{e.g.} \cite{Charron, Pedolsky, Regev}. We also find the energy conservation equation  \eqref{galEEuler}:
\begin{equation}
\partial_t\left(
\varrho\left( e+\frac{\mathbf{v}^2-\pmb{\omega}^2
\mathbf{x}^2+(\pmb{\omega}\cdot \mathbf{x})^2}{2}\right)
\right)
=-\text{\bf div}\, \pmb{\Pi}^{\text{G}},
\label{galEEuler-rot} 
\end{equation}
with 
\begin{equation}
\pmb{\Pi}^{\text{G}}=
\varrho\mathbf{v}\left( h+\frac{\mathbf{v}^2-\pmb{\omega}^2
\mathbf{x}^2+(\pmb{\omega}\cdot \mathbf{x})^2}{2}\right)
-\kappa^{\text{G}} \, \text{\bf grad}\, T-\mathbf{v}\cdot\pmb{\Sigma}^{\text{G}}_{(1)}
\label{galEPi-rot} 
\end{equation}
and 
\begin{equation}
\Sigma^{\text{G}}_{(1)ij}= \eta^{\text{G}}\left(\partial_i v_j+\partial_j v_i\right)+\left(\zeta^{\text{G}}-\frac{2}{3}\eta^{\text{G}}\right)\delta_{ij}\partial_k v^k.
\label{galsig-1-rot} 
\end{equation}
Alternatively, using \eqref{galEbis}, the energy equation reads:
\begin{equation}
\varrho\dfrac{\text{d}}{\text{d}t}\left(e+\frac{\mathbf{v}^2-\pmb{\omega}^2
\mathbf{x}^2+(\pmb{\omega}\cdot \mathbf{x})^2}{2}\right)=-\textbf{div}p\mathbf{v}+\kappa^{\text{G}} \Delta T+\textbf{div}\left(\mathbf{v}\cdot\pmb{\Sigma}^{\text{G}}_{(1)}\right).
\end{equation}
The temporal variation of the total energy per mass is given by the divergences of the pressure, the thermal conduction and the viscous stress fluxes.

\subsubsection*{Inflating space}  

The dynamics of a non-perfect fluid moving on an inflating space can be studied considering: 
\begin{equation}
a_{ij}(t,\mathbf{x})=\exp\left(\alpha(t)\right)\tilde{a}_{ij}(\mathbf{x}),\quad \Omega = 1,\quad \mathbf{w}=0.
\end{equation}
The space dimension $d$ is arbitrary here, therefore:
\begin{equation}
\ln\sqrt{a}=d\,\frac{\alpha}{2}+\ln \sqrt{\tilde{a}}.
\end{equation}
The fluid equations obtained from  \eqref{galC}, \eqref{galEbis} and \eqref{galMEuler} become 
\begin{eqnarray}
&& \partial_t\varrho+\dfrac{\alpha' }{2}d\varrho+\textbf{div}\varrho\mathbf{v}=0, \label{galCinfl}\\
&& \varrho\frac{\text{d}}{\text{d}t}\left(e+\frac{\mathbf{v}^2}{2}\right)+\dfrac{\alpha'}{2} \left(\varrho \mathbf{v}^2
+dp-\text{Tr}\, \pmb{\Sigma}^{\text{G}}
\right) + \textbf{div}\left(p\mathbf{v}+\pmb{Q}^\text{G}-\mathbf{v}\cdot \pmb{\Sigma}^{\text{G}}\right)=0
, \label{galEinfl}\\
\label{galMinfl}
&&\varrho \dfrac{\text{d}v^i}{\text{d}t}+\alpha'\varrho v^i+\nabla^i p -\nabla_j\Sigma^{\text{G}ij}=0.
\end{eqnarray}
where $\alpha' = \nicefrac{\text{d}\alpha}{\text{d}t}$ and 
$\text{Tr}\, \pmb{\Sigma}^{\text{G}}=a^{ij}\Sigma^{\text{G}}_{\hphantom{\text{G}}ij}$.

The continuity equation \eqref{galCinfl} has an extra term proportional to $\varrho$. This reflects the  change of density due to $\alpha'$. For a static fluid one finds the familiar result $\varrho=\varrho_0 \text{e}^{-\nicefrac{d\alpha }{2}}$: for a space expanding in time, the density is getting diluted. In Euler's equation \eqref{galMinfl}, a similar term creates a force proportional to the velocity field. For positive $\alpha'$, time dependence acts effectively like a friction. A similar conclusion is drawn from the energy conservation equation  \eqref{galEinfl}.

\subsection{Two-dimensional Carrollian fluids and the Robinson--Trautman dynamics} \label{EXART}

Consider now a two-dimensional surface $\mathscr{S}$, endowed with a complex chart $(\zeta, \bar\zeta)$ and a time-dependent metric of the form
\begin{equation}
\label{RTbdymetspace}
\text{d}\ell^2=\frac{2}{P(t,\zeta,\bar\zeta)^2}\text{d}\zeta\text{d}\bar\zeta.
\end{equation}
In this case the Carrollian shear $\pmb{\xi}^{\text{C}}$ \eqref{carsh} vanishes.
We assume that the Carrollian frame has $\pmb{b}=0$ and $\Omega=1$, and that the Carrollian kinematical variable $\pmb{\beta}$ also vanishes. Hence, the Carrollian inertial acceleration $\pmb{\varphi}$ \eqref{caracc} and inertial vorticity $\pmb{\varpi}$ \eqref{carom} vanish together with the kinematical acceleration $\pmb{\gamma}^{\text{C}}$ \eqref{carkinacc} and kinematical vorticity $\pmb{w}$ \eqref{carkinvort}. We further assume that $\pmb{\pi}$ and $\pmb{\Xi}$ vanish, so that the friction and heat-transport phenomena are captured exclusively by $ \pmb{Q}^\text{C}$ and $ \pmb{\Sigma}^{\text{C}}$. Hence $e_{\text{e}}=\varepsilon$, $\Pi^{\text{C}}_{\hphantom{\text{C}}j}=Q^{\text{C}}_{\hphantom{\text{C}}j}$ and 
$\Pi^{\text{C}}_{\hphantom{\text{C}}ij}=p a_{ij}$.

We will here study a conformal Carrollian fluid. In this case (see \eqref{con-carroll}), the Gibbs--Duhem equation reads
\begin{equation}
 \varepsilon(t,\zeta, \bar\zeta) = 2 p(t,\zeta, \bar\zeta),
 \label{conf}
\end{equation} 
and the viscous tensor is traceless:
\begin{equation}
\label{carfsd-tracefree} 
\Sigma^{\text{C}\zeta\bar\zeta}=0.
\end{equation}
The generic set of equations of motion for the Carrollian fluid at hand is  (see \eqref{carEcon}, \eqref{carGcon}, \eqref{carHcon})
\begin{eqnarray}
&&\mathcal{E}=
3\varepsilon\partial_t \ln P-\partial_t \varepsilon - \text{\bf div}\, \pmb{Q}^\text{C}=0,
 \label{carE2d} 
\\
&&\pmb{\mathcal{G}}=
 \text{\bf grad}\,p= 0,
  \label{carG2d}\\
&&\pmb{\mathcal{H}}= \partial_t \pmb{Q}^\text{C}-2 \pmb{Q}^\text{C}\partial_t \ln P
-\text{\bf div}\, \pmb{\Sigma}^{\text{C}}=0,
 \label{carH2d} 
\end{eqnarray}
together with Eq.  \eqref{carFcon}, $\mathcal{F}=0$, identically satisfied due to the absence of shear.  Equations \eqref{carE2d}, \eqref{carG2d} and \eqref{carH2d} are covariant under Weyl transformations mapping $P(t,\zeta,\bar\zeta)$ onto $\mathcal{B}(t,\zeta,\bar\zeta)P(t,\zeta,\bar\zeta)$ with  $\mathcal{B}(t,\zeta,\bar\zeta)$ an arbitrary function. 
 
 The momentum equation \eqref{carG2d} states that the pressure $p$ is space-independent, which is not a surprise for a fluid at $\pmb{\beta}=0$ in a Carrollian frame with vanishing $\pmb{b}$ and constant $\Omega$. The same holds for the energy, due to the equation of state. 
 
 In order to proceed we must introduce some further assumptions regarding the heat current  
 and the viscous stress tensor. These quantities are rooted to the unknown microscopic properties of the Carrollian fluids. As already mentioned earlier in Sec. \ref{FLUID2}, due to the absence of motion even at a microscopic level, it  is tempting to assign a geometric rather than a statistical or kinetic origin to Carrollian thermodynamics. We may therefore define the \emph{Carrollian temperature} as
\begin{equation}
\label{kinT}
\kappa^{\text{C}}T(t,\zeta,\bar \zeta) = \left\langle \kappa^{\text{C}}T\right\rangle(t)+\kappa^{\prime}K(t,\zeta,\bar \zeta)
-\kappa^{\prime}\left\langle K \right\rangle(t),
\end{equation} 
where 
$K$ the Gaussian curvature of \eqref{RTbdymetspace}:
\begin{equation}
\label{K}
K=\Delta\ln P
\end{equation}
with $\Delta = 2P^2 \partial_{\bar\zeta} \partial_\zeta$ the ordinary two-dimensional Laplacian operator. 
The thermal conductivity $\kappa^{\text{C}}$ is not constant in general because the identification with the curvature scalar endows the product $\kappa^{\text{C}}T$ with a conformal weight 2, whereas the temperature $T$ has weight 1.  We also introduced a constant $\kappa^{\prime}$ for matching the dimensions. 
In expression \eqref{kinT}, $\left\langle \kappa^{\text{C}} T\right\rangle(t)$ is an \emph{a priori} arbitrary time-dependent reference temperature (times thermal conductivity), and the brackets are meant to average over $\mathscr{S}$:\footnote{Here  $\text{d}^2\zeta=-i\,\text{d}\zeta\wedge\text{d}\bar\zeta$. If $\mathscr{S}$ is non-compact a limiting procedure is required for defining the integrals.}
\begin{equation}
\langle
f 
\rangle(t)= \frac{1}{A}
\int_\mathscr{S}\frac{\text{d}^2\zeta}{P^2} f(t,\zeta,\bar \zeta), \quad A =\int_\mathscr{S}\frac{\text{d}^2\zeta}{P^2}.
\end{equation} 

Equipped with a temperature, we  define next the heat current as its gradient
\begin{equation}
\label{q}
\pmb{Q}^\text{C} = - \text{\bf grad}\, \kappa^{\text{C}} T= -\kappa^{\prime}\,  \text{\bf grad}\, K,
\end{equation} 
following first-order Carrollian hydrodynamics, Eq. \eqref{Car1storderq}. Here, we assume  this expression be exact, \emph{i.e.} without higher-derivative contributions.  With these definitions, the heat equation \eqref{carE2d} for the Carrollian fluid at hand reads:
\begin{equation}
\label{RT}
3\varepsilon \partial_t \ln P-\partial_t \varepsilon +\kappa^{\prime} \Delta K=0,
\end{equation} 
where we have used the equation of state \eqref{conf}. This is a dynamical equation for $P(t,\zeta,\bar \zeta)$, given $\varepsilon(t)$. Carrollian dynamics, within the framework set by our definitions of temperature and heat current, is therefore purely geometrical and describes the evolution of the hosting space $\mathscr{S}$ rather than the fluid itself. This is not a surprise because the fluid does not move.  Going in the Carrollian limit from a relativistic set-up, amounts to trading the dynamics of the fluid for that of the supporting geometry.

We must finally impose 
Eq. \eqref{carH2d}. As we mentioned in the general discussion of Sec. \ref{FLUID2}, this is not an evolution equation, but instead a constraint among the heat current, the viscous stress tensor and the ambient geometry. 
Thus, we can integrate it using \eqref{q}. We find
\begin{equation}
\label{pi}
\pmb{\Sigma}^{\text{C}}  = -\frac{2\kappa^{\prime}}{P^2}
\left(
\partial_{\zeta} \left(P^2\partial_t\partial_{\zeta}
\ln P
\right)\text{d}\zeta^2+ 
\partial_{\bar\zeta} \left(P^2\partial_t\partial_{\bar\zeta}
\ln P
\right)
\text{d}\bar\zeta^2 \right),
\end{equation}
up to a divergence-free, trace-free symmetric tensor. The viscous stress tensor for the Carrollian fluid at hand is therefore geometric, as is the heat current, and both appear as third-order derivatives of the metric.  Actually, 
the effective expansion generally defined for Carrollian fluids as in \eqref{carexp}, reads 
here:
\begin{equation}
\label{carRT} 
\theta^{\text{C}} =-2 \partial_t \ln P.
\end{equation}
It enables to view $\pmb{\Sigma}^{\text{C}}$  as a velocity third derivative through the writing
\begin{equation}
\label{sigmaoffdpr}
\Sigma^{\text{C}}_{\hphantom{\text{C}}ij}=\kappa^{\prime}\left(
\nabla_i \nabla_j  \theta^{\text{C}} -\frac{1}{2} a_{ij} \nabla^k \nabla_k  \theta^{\text{C}} \right).
\end{equation}
Notice that in the two-dimensional background under consideration \eqref{RTbdymetspace}, the viscous tensor $\pmb{\Sigma}^{\text{C}} $ could not have received an $\eta^{{\text{C}}}$-induced first-order derivative correction as in \eqref{Car1stordertau2} because the Carrollian shear $\xi^{\text{C}}_{ij}$ given in \eqref{carsiglim} vanishes here identically. However, since the Carrollian expansion $\theta^{\text{C}} $ is non-zero, the absence of first-order derivative correction \eqref{Car1stordertau2} 
implies that for the fluid at hand $\zeta^{{\text{C}}}=0$.

Equation \eqref{RT}, which is at the heart of two-dimensional conformal Carrollian fluid dynamics, is actually known as Robinson--Trautman. It emerges when solving four-dimensional Einstein equations, assuming the existence of a null, geodesic and shearless congruence \cite{RTorig}. In vacuum or in the presence of a cosmological constant, Goldberg--Sachs theorems state that the corresponding spacetime is algebraically
special and the whole dynamics boils down to the Robinson--Trautman equation with $\varepsilon (t) = 4\kappa^{\prime} M (t)$ and $\kappa^{\prime} = \nicefrac{1}{16\pi G}$ (using \eqref{K}):  
\begin{equation}
\label{RT2}
\Delta \Delta  \ln P+12M\partial_t \ln P-4\partial_t M =0.
\end{equation} 
In that framework, the time dependence of the mass function $M(t)$ can be reabsorbed by an appropriate coordinate transformation (see \emph{e.g.} \cite{GP}) and Robinson--Trautman equation becomes then  
\begin{equation}
\label{RT-time-ind}
2 \partial_{\bar\zeta} \partial_\zeta P^2 \partial_{\bar\zeta} \partial_\zeta \ln P=3M \partial_t\left( \frac{1}{P^2}\right)
\end{equation}
with $M$ constant related to the Bondi mass. This is a parabolic equation describing a Calabi flow on a two-surface \cite{Tod:1989}.  

The reason why Robinson--Trautman appears both as a heat equation in conformal Carrollian fluids and as a remnant of four-dimensional Einstein equations is the holographic relationship between gravity and fluid dynamics. The two-dimensional conformal Carrollian fluid studied here originates from flat Robinson--Trautman spacetime holography \cite{CMPPS2}. Similarly 
Robinson--Trautman equation is the heat equation for $2+1$-dimensional relativistic boundary fluids emerging holographically from four-dimensional anti-de Sitter Robinson--Trautman spacetimes \cite{Ciambelli:2017wou}.

\section{Conclusions}

We can summarize our method and results as follows. 

A general relativistic spacetime metric is covariant under diffeomorphisms. When  put in Zermelo form, the data $\Omega(t)$, $w^i(t,\mathbf{x})$ and $a_{ij}(t,\mathbf{x})$ transform under Galilean diffeomorphisms $t'=t'(t)$ and $\textbf{x}^{\prime}=\textbf{x}^{\prime}(t, \textbf{x})$ as they should to comply with the infinite-$c$ non-relativistic expectations. This observation is made by analyzing the relativistic particle motion and its classical limit. It provides the appropriate framework for studying the general non-relativistic Galilean fluid dynamics as an infinite-$c$ limit of the relativistic one. In this manner, we have obtained the general equations \emph{i.e.} continuity, energy-conservation and Euler, valid on any spatial background, potentially time-dependent, and observed from an arbitrary frame.  These equations transform covariantly under Galilean diffeomorphisms. 

Alternatively, one can study relativistic instantonic space-filling branes and the small-$c$ behaviour of their dynamics. The latter is invariant under Carrollian diffeomorphisms $t'=t'(t, \textbf{x})$ and $\textbf{x}^{\prime}=\textbf{x}^{\prime}(\textbf{x})$, and Randers--Papapetrou form  is the best designed spacetime metric because the data  $\Omega(t,\mathbf{x})$, $b_i(t,\mathbf{x})$ and $a_{ij}(t,\mathbf{x})$ transform as expected from the non-relativistic limit (which is actually ultra-relativistic). In Randers--Papapetrou backgrounds one can study relativistic fluids and their Carrollian limit at vanishing velocity of light. This limit exhibits a new connection, which naturally fits into the emerging Carrollian geometry. One obtains in this way the general equations for the Carrollian fluids, manifestly covariant under Carrollian diffeomorphisms. 

Several comments are in order here. 

{ The Carrollian set we have reached is made of two scalar and two vector equations. The first scalar is an energy conservation, whereas the first vector is a momentum conservation. As there is no motion (due to $c=0$), there is no velocity field. Nonetheless there is a kinematical fluid variable (an ``inverse velocity'') accompanied by the pressure and energy density, related through an equation of state. We also find two heat currents and two viscous stress tensors. 
The Carrollian-fluid data cannot be naturally encapsulated all together in an energy--momentum tensor or an energy flux, as it happens in the Galilean case. Half of the equations concern exclusively the heat currents and the viscous stress tensors, relating them intimately to the ambient geometry and the Carrollian frame. We should stress here that we have made a specific assumption on the small-$c$ behaviour of the relativistic viscous stress tensor and  heat current, or equivalently of the transport coefficients. The number and the structure of the equations finally obtained reflects this unavoidable ansatz, inspired from the holographic Carrollian fluids met in flat-space gravity/fluid correspondence \cite{CMPPS2}.\footnote{Concrete examples of exact Carrollian fluids are described in this reference.} Going further in understanding this ansatz, and the physics behind the equations of motion, would require a microscopic analysis of Carrollian fluids. }

Despite the absence of velocity field in Carrollian hydrodynamics, the concept of derivative expansion still  holds. At each order one can define covariant tensors build on time and space derivatives of $a_{ij}$, $b_i$ and $\beta_i$, as we met at first order with the shear and the expansion. The heat current and the viscous stress tensor can be expanded in these tensors, introducing phenomenological  transport coefficients of increasing order.

Regarding Carrollian hydrodynamics, one could exploit a radically different perspective. Instead of defining a Carrollian fluid as the zero-$c$ limit of a relativistic fluid in some Randers--Papapetrou background, one could simply try to build a fluid-like -- \emph{i.e.} continuous -- generalization of an instantonic $d$-brane, directly within a Carrollian structure. This would promote the ``inverse velocity'' $\partial_i t$ of the elementary $d$-brane described by $t=t(\mathbf{x})$ into an ``inverse velocity field'' reminiscent of $\beta_ i+b_i$ and transforming as in \eqref{cardift} under a Carrollian diffeomorphism. This could be the starting point for designing the dynamics of this new continuous Carrollian medium. Irrespective of the viewpoint chosen for describing Carrollian continuous media, zero-$c$ limit of ordinary relativistic fluids or $d$-brane continuums, a great deal of fundamental thermodynamics, kinetic theory, derivative expansions, equilibrium and transport dynamics remains to be unravelled.

In conclusion of our general work, we have presented some examples. Those on Galilean hydrodynamics illustrate the power of the formalism for handling general, time-dependent and curved host spaces, potentially observed from non-inertial frames. The example of two-dimensional Carrollian fluid is interesting because it introduces the concept of geometric temperature and treats dissipative phenomena exactly \emph{i.e.} by solving explicitly all the equations but one, finally brought in the canonical form of a Calabi flow on the two-dimensional surface. The Carrollian fluid dynamics translates into a dynamics for the geometry. This example has important implications in asymptotically flat holography \cite{CMPPS2} of Robinson--Trautman spacetimes. 

\section*{Acknowledgements}

We would like to thank G. Bossard for sharing views on Carrollian dynamics, and A. Restuccia for his interest in the general Galilean fluid equations. 
We thank each others home institutions for hospitality and financial support. 
This work was supported by the ANR-16-CE31-0004 contract \textsl{Black-dS-String}.

\appendix 

\section{Christoffel symbols, transformations and connections}

We provide here a toolbox for working out the Galilean and Carrollian limits in the Zermelo and Randers--Papapetrou backgrounds, and checking the covariance properties of the set of equations reached by this method. These properties are bound to the emergence of novel Galilean and Carrollian connections, and covariant derivatives, which are discussed together with the associated curvature tensors. In the Carrollian case, an extra conformal connection is also  presented, relevant when studying conformal Carrollian fluids.

\subsection{Zermelo metric}  \label{zerapp}

\subsubsection*{Christoffel symbols}

The Zermelo metric \eqref{galzerm} has components (in the coframe $\left\{\text{d}x^0=c\text{d}t,\text{d}x^i\right\}$):
 \begin{equation}
 \label{compzerm} 
g^\text{Z}_{\mu\nu}\to
\begin{pmatrix}   -\Omega^2+\frac{\mathbf{w}^2}{c^2} &-\frac{w_k}{c}\\  
-\frac{w_i}{c} & a_{ik}
 \end{pmatrix} ,\quad
 g^{\text{Z}\mu\nu}\to
 \frac{1}{\Omega^2}
 \begin{pmatrix}   -1 &-\frac{w^j}{c}\\  
-\frac{w^i}{c} & \Omega^2 a^{ij}-\frac{w^i w^j}{c^2}
 \end{pmatrix},
\end{equation}
where $w_k=a_{kj}w^j$. Its determinant reads:
 \begin{equation}
 \label{detzerm} 
\sqrt{-g}=\Omega \sqrt{a},
\end{equation}
where $a$ is the determinant of $a_{ij}$.  We remind that $\Omega$ depends on time only, whereas 
 $a_{ij}$ and  $w_{i}$ also depend on space.

The Christoffel symbols are easily computed. We are interested in their large-$c$ behaviour for which one obtains the following:
\begin{eqnarray}
\Gamma^0_{00}&=&
\frac{1}{c}\partial_t \ln \Omega 
++\frac{w^i}{2c^3\Omega^2}\left(
\partial_i \mathbf{w}^2+w^j\partial_t a_{ij}
\right)
+\text{O}\left(\nicefrac{1}{c^5}\right),\\
\Gamma^0_{0i}&=&
-\frac{1}{2c^2\Omega^2}\left(
w_j\partial_i w^j
+w^j\partial_j w_i
+w^j\partial_t a_{ij}\right)
+\text{O}\left(\nicefrac{1}{c^4}\right), \\
\Gamma^0_{ij}&=&
\frac{1}{c\Omega^2}\left(\frac{1}{2}\left(
\partial_i w_j
+\partial_j w_i
+\partial_t a_{ij}\right)
-w_k \gamma^k_{ij}
\right),
\\
\Gamma^i_{00}&=&\dfrac{1}{c^2}\left(w^i \partial_t \ln \Omega-a^{ik}\left(\partial_t w_k+\partial_k \dfrac{\mathbf{w}^2}{2}\right)\right)+\text{O}\left(\nicefrac{1}{c^4}\right),\\
\Gamma^i_{j0}&=&\dfrac{a^{ik}}{2c}\left(\partial_k w_j-\partial_j w_k+\partial_t a_{jk} \right)+\text{O}\left(\nicefrac{1}{c^3}\right),\\
\Gamma^i_{jk}&=&\gamma^i_{jk}+\text{O}\left(\nicefrac{1}{c^2}\right),
\end{eqnarray}
where 
\begin{equation}
\label{dgamma}
\gamma^i_{jk}=\dfrac{a^{il}}{2}\left(\partial_j a_{lk}+\partial_k a_{lj}-\partial_l a_{jk}\right)
\end{equation}
are the Christoffel symbols for the $d$-dimensional metric $a_{ij}$. Note also
\begin{equation}
\Gamma^\mu_{\mu0}=\frac{1}{c}\partial_t \ln \left(\sqrt{a}\Omega\right),\quad
\Gamma^\mu_{\mu i}=\partial_i \ln \sqrt{a}.
\end{equation}
With these data it is possible to compute the divergence of the fluid energy--momentum tensor \eqref{conTgalexp0} and \eqref{conTgalexpi}.

\subsubsection*{Covariance}

In order to check the covariance \eqref{galCEdif} and  \eqref{galMdif},
\begin{equation}
\mathcal{C}^\prime= \mathcal{C},\quad \mathcal{E}^\prime=\mathcal{E}\quad
\mathcal{M}^\prime_i = J^{-1l}_{\hphantom{-1}i} \mathcal{M}_l,
\nonumber
\end{equation}
for the Galilean fluid dynamics under Galilean diffeomorphisms  \eqref{galdifs} 
\begin{equation}
\nonumber
t'=t'(t)\quad \text{and} \quad \textbf{x}^{\prime}=\textbf{x}^{\prime}(t, \textbf{x}),
\end{equation}
with Jacobian functions  \eqref{galj}
\begin{equation}
\nonumber
J(t)=\frac{\partial t'}{\partial t},\quad j^i(t,\mathbf{x}) = \frac{\partial x^{i\prime}}{\partial t},\quad 
J^i_j(t,\mathbf{x}) = \frac{\partial x^{i\prime}}{\partial x^{j}},
\end{equation}
we can use several simple covariant blocks. We first remind 
\eqref{galdifa}, \eqref{galdifv}, \eqref{galdifw}, \eqref{galdifom}: 
\begin{equation}
\nonumber
a^{\prime}_{ij} =a_{kl} J^{-1k}_{\hphantom{-1}i} J^{-1l}_{\hphantom{-1}j} ,\quad
v^{\prime k}=\frac{1}{J}\left(J^k_i v^i+j^k\right),\quad
w^{\prime k}=\frac{1}{J}\left(J^k_i w^i+j^k\right),\quad
\Omega^{\prime }=\frac{\Omega}{J},
\end{equation}
implying in particular
\begin{equation}
v^{\prime}_{k}=\frac{J^{-1i}_{\hphantom{-1}k}}{J}\left(v_i+a_{ij}J^{-1j}_{\hphantom{-1}l}j^l\right),\quad
w^{\prime}_{k}=\frac{J^{-1i}_{\hphantom{-1}k}}{J}\left(w_i+a_{ij}J^{-1j}_{\hphantom{-1}l}j^l\right)\end{equation}
with
\begin{eqnarray}
\label{galdelt}
\partial^\prime_t&=&\frac{1}{J}\left(\partial_t-j^kJ^{-1i}_{\hphantom{-1}k}\partial_i\right),\\
\label{galdelj}
\partial^\prime_j&=&J^{-1i}_{\hphantom{-1}j}\partial_i.
\end{eqnarray}

Consider now $A^k$ and $B^k$, the components of fields transforming like $v^k$ or  $w^k$ (gauge-like transformation) and  $V^k$ a field transforming like $\frac{v^k-w^k}{\Omega}$ \emph{i.e.} like a genuine vector:
\begin{equation}
\label{gal-trans1}
A^{\prime k}=\frac{1}{J}\left(J^k_i A^i+j^k\right),\quad B^{\prime k}=\frac{1}{J}\left(J^k_i B^i+j^k\right),\quad
V^{\prime k}=J^k_i V^i.
\end{equation}
Consider also a scalar and a rank-two  tensor
\begin{equation}
\label{gal-trans2}
\Phi^\prime=\Phi,\quad
S^{\prime}_{ij} =S_{kl} J^{-1k}_{\hphantom{-1}i} J^{-1l}_{\hphantom{-1}j}.
\end{equation}
The basic transformation rules are as follows:
 \begin{eqnarray}
  \label{gal-1}
\dfrac{A^{\prime k}-B^{\prime k}}{\Omega^\prime}
&=&J^k_i \dfrac{A^{i}-B^{i}}{\Omega},
\\
 \label{gal1}
 \frac{1}{\sqrt{a^\prime}}\partial^\prime_t\left( \sqrt{a^\prime}\Phi^\prime\right)+\nabla^\prime_i\left(\Phi^\prime A^{\prime i}\right)&=&
\frac{1}{J}\left(  \frac{1}{\sqrt{a}}\partial_t\left(\sqrt{a} \Phi \right)+\nabla_ i\left(\Phi A^i\right)\right),
  \\ \label{gal2}
\nabla^\prime_iV^{\prime i}&=&\nabla_ iV^i,
\\  \label{gal3}
\nabla^\prime_{(i}A^\prime_{j)}+\frac{1}{2}\partial^\prime_ta^\prime_{ij} &=& \frac{1}{J}\left( 
\nabla_{(k}A_{l)}+\frac{1}{2}\partial_ta_{kl} 
\right)J^{-1k}_{\hphantom{-1}i} J^{-1l}_{\hphantom{-1}j},
\\  \label{gal3bis}
\nabla^{\prime(i}A^{\prime j)}-\frac{1}{2}\partial^\prime_ta^{\prime ij} &=& \frac{1}{J}\left( 
\nabla^{(k}A^{l)}-\frac{1}{2}\partial_ta^{kl} 
\right)J_{k}^{i} J_{l}^{j},
\\  \label{gal4}
\nabla^\prime_iS^{\prime ij}&=&J^{j}_{l}\nabla_ i S^{il},
\\ \label{gal5}
\frac{1}{\Omega^\prime} \left(\partial^\prime_tV^\prime_{i}+A^{\prime j}\nabla^\prime_jV^\prime_{i} +V^\prime_{j}\nabla^\prime_i B^{\prime j}\right)
 &=& 
\frac{J^{-1k}_{\hphantom{-1}i}}{\Omega}\left(  \partial_tV_{k}+A^{j}\nabla_jV_{k} +V_{j}\nabla_k B^{ j}
\right),
\\
 \label{gal6}
\Delta^\prime A^\prime_i +r_ i^{\prime m}A^\prime_m
+a^\prime_{ik} a^{\prime mn}\partial^\prime_t\gamma^{\prime k}_{\hphantom{\prime}mn}
&=&
\frac{J^{-1j}_{\hphantom{-1}i}}{J}
\left(\Delta A_j+r_ j^{\hphantom{i}m}A_m
+a_{jk} a^{mn}\partial_t\gamma^k_{mn}\right).
\end{eqnarray}
In the above expressions, $\nabla_ i$, $\Delta$ and $r_ {ij}$ are associated with the $d$-dimensional Levi--Civita connection $\gamma^i_{jk}$ displayed in \eqref{dgamma}.

As a final comment regarding Galilean covariance properties, we would like to stress that the action of $\partial_t$ spoils the transformation rules displayed in \eqref{gal-trans1} and \eqref{gal-trans2}. This is both due to the transformation property of the partial time derivative \eqref{galdelt}, and to the time dependence of the Jacobian matrix $J^{i}_{j}$. A Galilean covariant time-derivative can be introduced, acting as follows on a vector:\footnote{For a detailed and general presentation of Galilean affine connections see
\cite{Bekaert:2014bwa, Bekaert:2015xua}.}
\begin{equation}
\label{galfcovder} 
\frac{1}{\Omega}\frac{\text{D}V^i}{\text{d}t}= \frac{1}{\Omega}\left[\left(\partial_t +v^j\nabla_j\right)V^i-V^j \nabla_j w^i\right]=
\frac{1}{\Omega}\frac{\text{d}V^i}{\text{d}t}-\frac{1}{\Omega}V^j \nabla_j w^i
,
\end{equation}
and resulting in a genuine vector under Galilean diffeomorphisms.
Here, the frame velocity $w^k$ plays the r\^ole of a connection, and the Galilean covariant time-derivative generalizes the material derivative $\nicefrac{\text{d}}{\text{d}t}$ introduced in \eqref{galfder}. The latter is covariant only when acting on scalar functions $f$, hence we set $\frac{\text{D}f}{\text{d}t}=\frac{\text{d}f}{\text{d}t}$. Expression \eqref{galfcovder} is easily extended for tensors of arbitrary rank using the Leibniz rule,
as \emph{e.g.} for one-forms:
\begin{equation}
\label{galfcovderf} 
\frac{1}{\Omega}\frac{\text{D}V_i}{\text{d}t}=\frac{1}{\Omega}\frac{\text{d}V_i}{\text{d}t}+\frac{1}{\Omega}V_j\nabla_i w^j.
\end{equation}
Notice that the Galilean covariant time-derivative at hand is not ``metric compatible'':
\begin{equation}
\label{galfcovdermet} 
\frac{1}{\Omega}\frac{\text{D}a_{ij}}{\text{d}t}=\frac{1}{\Omega}\left(\partial_ta_{ij}+2\nabla_{(i}w_{j)}\right).
\end{equation}
This result is actually expected because a covariant time-derivative of the metric should be interpreted as an extrinsic curvature. Indeed, expression \eqref{galfcovdermet} divided by $2c$ 
is exactly identified with the spatial components $K_{ij}$ of constant-$t$ hypersurfaces extrinsic curvature in the Zermelo background \eqref{galzerm}, \eqref{compzerm}.
  
The commutator of covariant time and space derivatives reveals a new piece of curvature, which appears in Galilean geometries, on top of the standard Riemann tensor associated with the spatial covariant derivative $\nabla_i$. It is  
encapsulated in a one-form $\text{d}\theta^{\text{G}}$, as one observes from: 
\begin{equation}
\left[\frac{1}{\Omega}\frac{\text{D}}{\text{d}t},\nabla_i\right]V^i= 
V^i \partial_i \theta^{\text{G}}
+\nabla_j\left(V^i\nabla_i \left(\frac{w^j-v^j}{\Omega}\right)\right),
\label{galriemanntime}
\end{equation}
where $\theta^{\text{G}}$ is a scalar function introduced in \eqref{galexp} as the Galilean effective expansion: 
\begin{equation}
\nonumber
\theta^{\text{G}} = \frac{1}{\Omega}\left(
\partial_t \ln \sqrt{a}+\nabla_i v^i\right).
\end{equation}
This extra piece of curvature should not come as a surprise. It is a Galilean remnant of some ordinary components of Riemannian curvature in the original Zermelo spacetime.

\subsection{Randers--Papapetrou metric}  \label{rpapp}

\subsubsection*{Christoffel symbols}
The Randers--Papapetrou metric \eqref{carrp} has components (in the coframe $\left\{\text{d}x^0=c\text{d}t,\text{d}x^i\right\}$):
 \begin{equation}
 \label{comprp} 
g^\text{RP}_{\mu\nu}\to
 \begin{pmatrix}   -\Omega^2 &c \Omega b_j\\  
c \Omega b_i & a_{ij}-c^2 b_i b_j
 \end{pmatrix}
 ,\quad
 g^{\text{RP}\mu\nu}\to\frac{1}{\Omega^2}
  \begin{pmatrix}   -1 +c^2\pmb{b}^2 &c \Omega b^k \\  
c \Omega b^i &\Omega^2 a^{ik}
 \end{pmatrix}
,
\end{equation}
where $b^k=a^{kj}b_j$. The metric determinant is again given in  \eqref{detzerm}:
 \begin{equation}
 \label{detRP} 
\sqrt{-g}=\Omega \sqrt{a}.
\end{equation}
Here, $\Omega$,
 $a_{ij}$ and  $b_{i}$ depend on time $t$ and space $\mathbf{x}$.

The Christoffel symbols are  computed exactly in the present case:
\begin{eqnarray}
\label{rp-chris-first}
\Gamma^0_{00}&=&
\frac{1}{c}\partial_t \ln \Omega 
+c\left(b^i\partial_i \Omega
+\frac{1}{2}\left(\partial_t\pmb{b}^2-b_ib_j\partial_t a^{ij}
\right)\right),
\\
\Gamma^0_{0i}&=&
\left(1-\frac{1}{2}c^2\pmb{b}^2 \right)\partial_i \ln \Omega
+\frac{1}{2}c^2b^j\left(\partial_i b_j -\partial_j b_i-b_i\partial_j  \ln \Omega \right)\nonumber\\
&&+\frac{1}{2\Omega} b^j \partial_t\left(a_{ij}-c^2 b_i b_j\right),
\\
\Gamma^0_{ij}&=&-\frac{c}{2\Omega}\left(
\partial_ib_j
+
\partial_jb_i+c^2b^k
\left(
b_i\left(
\partial_jb_k
-\partial_k b_j\right)
+
b_j\left(
\partial_ib_k
-\partial_k b_i\right)
\right)
\right)
\nonumber
\\&&+\frac{cb_k}{\Omega}\gamma^k_{ij}+\frac{1-c^2\pmb{b}^2}{2\Omega^2}\left(\frac{1}{c}\partial_t a_{ij}-cb_j\left(\partial_t b_i+\partial_i\Omega\right)-cb_i\left(\partial_t b_j+\partial_j\Omega\right)
\right),
\\
\Gamma^i_{00}&=&\Omega a^{ij}\left(\partial_t b_j + \partial_j\Omega\right),
\\
\Gamma^i_{j0}&=&\frac{1}{2c}a^{ik}\left(
\partial_t\left(a_{kj}-c^2b_kb_j\right)+c^2\Omega \left(\partial_j b_k-\partial_k b_j\right)
-c^2\left(b_k \partial_j \Omega+b_j\partial_k \Omega\right)
\right),
\\
\Gamma^i_{jk}&=& \frac{c^2}{2}\left(\frac{b^i}{\Omega}\left(
b_j\left(
\partial_tb_k
+\partial_k\Omega\right)
+
b_k\left(
\partial_tb_j
+\partial_j\Omega\right)\right)-a^{il}
\left(
b_j\left(
\partial_kb_l
-\partial_l b_k\right)
+
b_k\left(
\partial_jb_l
-\partial_l b_j\right)
\right)
\right)
\nonumber
\\&&+\gamma^i_{jk}-\frac{b^i}{2\Omega}\partial_t a_{jk}
,
\label{rp-chris-last}
\end{eqnarray}
where $\gamma^k_{ij}$ are the $d$-dimensional 
Christoffel symbols: 
\begin{equation}
\label{dgammaC}
\gamma^i_{jk}=\dfrac{a^{il}}{2}\left(\partial_j a_{lk}+\partial_k a_{lj}-\partial_l a_{jk}\right).
\end{equation}
Note also
\begin{equation}
\Gamma^\mu_{\mu0}=\frac{1}{c}\partial_t \ln \left(\sqrt{a}\Omega\right),\quad
\Gamma^\mu_{\mu i}=\partial_i \ln \left(\sqrt{a}\Omega\right).
\end{equation}
With these data it is possible to compute the divergence of the fluid energy--momentum tensor \eqref{conTcarexp0} and \eqref{conTcarexpi}.

\subsubsection*{Covariance and the Levi--Civita--Carroll connection}

In order to check the covariance \eqref{carEFdif} and  \eqref{carvdif},
\begin{equation}
\mathcal{E}^\prime=  \mathcal{E},\quad \mathcal{F}^\prime= \mathcal{F},\quad
\mathcal{G}^{\prime i} = J^{i}_{j} \mathcal{G}^j,\quad\mathcal{H}^{\prime i} = J^{i}_{j} \mathcal{H}^j
\nonumber
\end{equation}
for the Carrollian fluid dynamics under Carrollian diffeomorphisms  \eqref{cardifs} 
\begin{equation}
\nonumber
t'=t'(t,\textbf{x})\quad \text{and} \quad \textbf{x}^{\prime}=\textbf{x}^{\prime}(\textbf{x}),
\end{equation}
with Jacobian functions  \eqref{carj}
\begin{equation}
\nonumber
J(t,\mathbf{x})=\frac{\partial t'}{\partial t},\quad j_i(t,\mathbf{x}) = \frac{\partial  t'}{\partial x^{i}},\quad 
J^i_j(\mathbf{x}) = \frac{\partial x^{i\prime}}{\partial x^{j}},
\end{equation}
we can use several simple covariant blocks. We first remind \eqref{galdifom},
\eqref{cardifa}, \eqref{cardifb}: 
\begin{equation}
\nonumber
a^{\prime}_{ij} =a_{kl} J^{-1k}_{\hphantom{-1}i} J^{-1l}_{\hphantom{-1}j} ,\quad
b^{\prime}_{k}=\left( b_i+\frac{\Omega}{J} j_i\right)J^{-1i}_{\hphantom{-1}k},\quad
\Omega^{\prime }=\frac{\Omega}{J},\quad
\end{equation}
and
\begin{eqnarray}
\label{cart}
\partial^\prime_t&=&\frac{1}{J}\partial_t,\\
\label{carjj}
\partial^\prime_j&=&J^{-1i}_{\hphantom{-1}j}\left(\partial_i-
\frac{j_i}{J}\partial_t\right).
\end{eqnarray}

From the above transformation rules we obtains:
\begin{eqnarray}
\label{cara1}
\frac{1}{\Omega^\prime}\partial^\prime_t  a^\prime_{ij}&=&\frac{1}{\Omega}\partial_t  a_{kl}J^{-1k}_{\hphantom{-1}i} J^{-1l}_{\hphantom{-1}j},\\
\label{cara2}
\frac{1}{\Omega^\prime}\partial^\prime_t \ln  \sqrt{a^\prime}&=&\frac{1}{\Omega}\partial_t  \ln  \sqrt{a},
\\
\label{carb1}
\partial^\prime_t  b^\prime_i+\partial^\prime_i \Omega^\prime&=&\frac{1}{J}
J^{-1j}_{\hphantom{-1}i} \left(
\partial_t  b_j+\partial_j \Omega\right),\\
\label{carb2}
\hat\partial_i^\prime &=&
J^{-1j}_{\hphantom{-1}i} \hat\partial_j,
\end{eqnarray}
where we have defined
\begin{equation}
\label{dhat}
\hat\partial_i=\partial_i+\frac{b_i}{\Omega}\partial_t.
\end{equation}

In view of the basic rules  \eqref{cart}, \eqref{carjj} and \eqref{cara1}--\eqref{carb2}, it is tempting to introduce a new connection for Carrollian geometry that we will call \emph{Levi--Civita--Carroll}, whose coefficients will be generalizations of the Christoffel symbols \eqref{dgammaC}:
\begin{equation}
\label{dgammaCar}
\begin{array}{rcl}
\hat\gamma^i_{jk}&=&\dfrac{a^{il}}{2}\left(
\hat\partial_j
a_{lk}+\hat\partial_k  a_{lj}-
\hat\partial_l a_{jk}\right)
\crbig
&=&
\dfrac{a^{il}}{2}\left(
\left(\partial_j +\frac{b_j}{\Omega}
\partial_t
\right)
a_{lk}+\left(\partial_k +\frac{b_k}{\Omega}
\partial_t
\right) a_{lj}-\left(\partial_l +\frac{b_l}{\Omega}
\partial_t
\right) a_{jk}\right)
\crbig
&=&
\gamma^i_{jk}+c^i_{jk}
\end{array}
\end{equation}
with $\gamma^i_{jk}$ and $\hat\partial_i$ defined in \eqref{dgammaC} and \eqref{dhat}.
We will refer to those as \emph{Christoffel--Carroll} symbols. They transform under Carrollian diffeomorphisms as ordinary Christoffel symbols under ordinary diffeomorphisms:
\begin{equation}
\label{dgammaCartra}
\hat\gamma^{\prime k}_{\hphantom{\prime}ij}=
J^k_n 
J^{-1l}_{\hphantom{-1}i} 
J^{-1m}_{\hphantom{-1}j}
\hat\gamma^n_{lm}
- 
J^{-1l}_{\hphantom{-1}i} 
J^{-1n}_{\hphantom{-1}j}
\partial_l
J^{k}_{n}.
\end{equation}
The emergence of this new set of connection coefficients should not be a surprise. Indeed one readily  shows that
\begin{equation}
\label{dgammaCarlim}
h_i^{\hphantom{i}\mu}\Gamma^k_{\mu\nu}h^\nu_{\hphantom{\nu}j}=\hat\gamma^{k}_{ij},
\end{equation}
where $\Gamma^k_{\mu\nu}$ are the $d+1$-dimensional Randers--Papapetrou Christoffel symbols
\eqref{rp-chris-first}--\eqref{rp-chris-last}, and $h_\nu^{\hphantom{\nu}\mu}$ the projector orthogonal to $\text{u}=\nicefrac{\partial_t}{\Omega}$ (as in \eqref{relproj}, \eqref{carvel}).

The Levi--Civita--Carroll covariant derivative acts symbolically as 
\begin{equation}
\label{Carcovder}
\hat{ \pmb{\nabla}}=\hat{\pmb{\partial} }
+\hat{ \pmb{\gamma}}=
\pmb{\partial}+\frac{\pmb{b}}{\Omega}\partial_t+\pmb{\gamma}+\pmb{c}
=\pmb{\nabla}+\frac{\pmb{b}}{\Omega}\partial_t+\pmb{c}
.
\end{equation}
For example, consider $\Phi$, $V^k$ and $S_{kl} $, the components of a scalar, a vector, and rank-two symmetric  tensor:
\begin{equation}
\Phi^\prime=\Phi,\quad V^{\prime i} = J^i_j V^j,\quad
S^{\prime}_{ij} =S_{kl} J^{-1k}_{\hphantom{-1}i} J^{-1l}_{\hphantom{-1}j},
\end{equation} 
the action of this new covariant derivative is
\begin{eqnarray}
\hat\partial_i\Phi&=&\partial_i \Phi+\frac{b_i}{\Omega}\partial_t\Phi
\label{Carcovdersca}
,
\\
\hat\nabla_iV^j&=&\partial_i V^j+\frac{b_i}{\Omega}\partial_tV^j+\hat\gamma^j_{il}V^l\nonumber
\\
&=&
\nabla_iV^j+\frac{b_i}{\Omega}\partial_tV^j+c^j_{il}V^l
\label{Carcovdervec}
,
\\
\hat\nabla_iV^i&=&\frac{1}{\sqrt{a}}\hat\partial_i\left(\sqrt{a}V^i\right)
\\
\nonumber
\hat\nabla_iS_{jk}&=&\partial_i S_{jk}+\frac{b_i}{\Omega}\partial_tS_{jk}-\hat\gamma^l_{ij}S_{lk}-\hat\gamma^l_{ik}S_{jl}\\
&=&
\nabla_i S_{jk}+\frac{b_i}{\Omega}\partial_tS_{jk}-c^l_{ij}S_{lk}-c^l_{ik}S_{jl}.
\label{Carcovde2ten}
\end{eqnarray}
All these transform as genuine tensors, namely:
\begin{eqnarray}
\hat\partial^\prime_i\Phi^\prime&=&J^{-1j}_{\hphantom{-1}i}\hat\partial_j\Phi,\\
\hat\nabla^\prime_iV^{\prime j}&=&J^{-1k}_{\hphantom{-1}i}J^{j}_{l}\hat\nabla_kV^l,\\
\hat\nabla^\prime_iV^{\prime i}&=&\hat\nabla_iV^i,\\
\hat\nabla^\prime_iS^\prime_{jk}&=&J^{-1m}_{\hphantom{-1}i}J^{-1n}_{\hphantom{-1}j}J^{-1l}_{\hphantom{-1}k}\hat\nabla_mS_{nl}.
\end{eqnarray}
Further elementary transformation rules are as follows:
 \begin{equation}
 \frac{1}{\Omega^\prime}\partial^\prime_t\Phi^\prime=
\frac{1}{\Omega}\partial_t \Phi , \quad \frac{1}{\Omega^\prime}\partial^\prime_t V^{\prime i}=J^i_j
\frac{1}{\Omega}\partial_t V^j, \quad \frac{1}{\Omega^\prime}\partial^\prime_t S^{\prime ij}=J^i_k J^j_l
\frac{1}{\Omega}\partial_t S^{kl}, \label{time}
 \end{equation}
as well as 
 \begin{equation}
\nabla^\prime_i V^{\prime i}+ \frac{b^\prime_i}{\Omega^\prime\sqrt{a^\prime}}\partial^\prime_t\left(\sqrt{a^\prime}V^{\prime i} \right)=\hat\nabla^\prime_i V^{\prime i}= \hat\nabla_ i V^i= \nabla_ i V^i+\frac{b_i}{\Omega\sqrt{a}}\partial_t\left(\sqrt{a}V^i \right),
    \label{car2}
 \end{equation}
and
 \begin{eqnarray}
\nonumber
& \nabla^\prime_k S^{\prime ki}+ \frac{b^\prime_k}{\Omega^\prime\sqrt{a^\prime}}\left(\partial^\prime_t\left(\sqrt{a^\prime}S^{\prime ki} \right)-\sqrt{a^\prime}
 S^{\prime k}_{\hphantom{\prime k}j}
 \partial^\prime_t a^{\prime ij}
 \right)
 - \frac{b^{\prime i}}{2\Omega^\prime}S^{\prime kl}  \partial^\prime_t a^\prime_{kl}
=\hat\nabla^\prime_k S^{\prime ki}=&
\\
 \label{car3}
&=J^i_j\hat\nabla_k S^{ kj} =J^i_j\left( \nabla_k S^{ kj} +
 \frac{b_k}{\Omega\sqrt{a}}\left(\partial_t\left(S^{kj}\sqrt{a}\right) -\sqrt{a}
 S^{k}_{\hphantom{k}l}
 \partial_t a^{jl}
 \right)
 - \frac{b^{j}}{2\Omega}S^{kl}  \partial_t a_{kl}
\right).&
\end{eqnarray}

\subsubsection*{Curvature, effective torsion and further properties of the Levi--Civita--Carroll connection}

The Levi--Civita--Carroll connection is metric,
\begin{equation}
\label{carconmet}
\hat\nabla_ia_{jk}=0.
\end{equation}
Furthermore, the usual torsion tensor vanishes:\footnote{Discussions on Carrollian affine connections can be found \emph{e.g.} in \cite{Bekaert:2015xua, Duval:2014lpa,Hartong:2015xda}. In particular, Ref. \cite{Bekaert:2015xua} provides a general classification of connections with or without torsion.}
\begin{equation}
\label{carcontorsion}
\hat t^k_{\hphantom{k}ij}=2\hat\gamma^k_{[ij]}=0.
\end{equation}
However, the new ordinary (as opposed to covariant) derivatives $\hat\partial_i$ defined in \eqref{dhat} 
do not commute. Indeed, acting on any arbitrary function they lead to 
\begin{equation}
\label{carconcomderf}
\left[\hat\partial_i,\hat\partial_j\right]\Phi=
\frac{2}{\Omega}\varpi_{ij}\partial_t \Phi,
\end{equation}
where $\varpi_{ij}$ are the components of the Carrollian vorticity defined in \eqref{carom} (explicitly in  \eqref{caromlim}) using the Carrollian acceleration $\varphi_i$ \eqref{caracc}:
\begin{equation}
\label{caromacc}
\varpi_{ij}=\partial_{[i}b_{j]}+b_{[i}\varphi_{j]}, \quad \varphi_i=\dfrac{1}{\Omega}\left(\partial_t b_i+\partial_i \Omega\right).
\end{equation}
Therefore, the Levi--Civita--Carroll connection has an \emph{effective torsion} as one can see from
\begin{equation}
\label{carcontor}
\left[\hat\nabla_i,\hat\nabla_j\right]\Phi=
\varpi_{ij}\frac{2}{\Omega}\partial_t\Phi,
\end{equation}
where $\Phi$ is a scalar. 

Similarly, one can compute the commutator of the Levi--Civita--Carroll covariant derivatives acting on a vector field:
\begin{equation}
\begin{array}{rcl}
\left[\hat\nabla_k,\hat\nabla_l\right]V^i&=&\left(
\hat\partial_k\hat\gamma^i_{lj}
-\hat\partial_l\hat\gamma^i_{kj}
+\hat\gamma^i_{km}\hat\gamma^m_{lj}
-\hat\gamma^i_{lm}\hat\gamma^m_{kj}
\right)V^j+\left[\hat\partial_k,\hat\partial_l\right]V^i
\\
&=& \hat r^i_{\hphantom{i}jkl}V^j+
\varpi_{kl}\frac{2}{\Omega}\partial_{t}V^i.
\end{array}
\label{carriemann}
\end{equation}
In this expression we have defined $ \hat r^i_{\hphantom{i}jkl}$, which are by construction components of a genuine tensor under Carrollian diffeomorphisms in $d$ dimensions. This should be called the \emph{Riemann--Carroll} tensor. It is made of several pieces, among which
$
\partial_k\gamma^i_{lj}
-\partial_l\gamma^i_{kj}
+\gamma^i_{km}\gamma^m_{lj}
-\gamma^i_{lm}\gamma^m_{kj},
$
which is \emph{not} covariant under Carrollian diffeomorphisms -- it is under ordinary $d$-dimensional diffeomorphisms though. The Ricci--Carroll tensor and the Carroll scalar curvature are thus
\begin{equation}
\label{carricci-scalar}
\hat r_{ij}=\hat r^k_{\hphantom{k}ikj},\quad \hat r=a^{ij}\hat r_{ij}.
\end{equation}
Notice that the Ricci--Carroll tensor is \emph{not} symmetric in general: $\hat r_{ij}\neq \hat r_{ji}$.

We would like to close this part with two remarks regarding Carrollian geometry and in particular Carrollian time. As readily seen in \eqref{time}, acting on any object tensorial under Carrollian diffeomorphisms, the time derivative $\partial_t$ provides another tensor.  For this reason, it was not necessary to define any ``temporal covariant derivative''. Our first remark is that the ordinary time derivative has an unsatisfactory feature: its action on the metric does not vanish. One is tempted therefore to
set a new time derivative  $\hat \partial_t$ such that
\begin{equation}
\label{cartimemet}
\hat \partial_ta_{jk}=0,
\end{equation}
while keeping the transformation rule under Carrollian diffeomorphisms:
\begin{equation}
\label{carcovdt}
{\hat\partial}^\prime_t=\frac{1}{J}\hat\partial_t.
\end{equation}
This is achieved by introducing a ``temporal Carrollian connection''
\begin{equation}
\label{dgammaCartime}
\hat\gamma^i_{\hphantom{i}j}=\frac{1}{2\Omega}a^{ik}\partial_t a_{kj}.
\end{equation}
Calling this a connection is actually inappropriate because it transforms as a genuine tensor under Carrollian diffeomorphisms:
\begin{equation}
\label{dgammaCartratime}
\hat\gamma^{\prime k}_{\hphantom{\prime k}j}=
J^k_n 
J^{-1m}_{\hphantom{-1}j}
\hat\gamma^n_{\hphantom{n}m}.
\end{equation}
In fact, the trace of this object is the Carrollian expansion introduced in 
\eqref{carexp1}:
\begin{equation}
\label{carexp-tempcon}
\theta^{\text{C}}=
\dfrac{1}{\Omega}              
\partial_t \ln\sqrt{a}=\hat\gamma^{i}_{\hphantom{i}i},
\end{equation}
whereas its traceless part is the Carrollian shear defined in \eqref{carsh}:
\begin{equation}
\label{carsh-tempcon}
\xi^{\text{C}i}_{\hphantom{\text{C}i}j}= \hat\gamma^i_{\hphantom{i}j}-\frac{1}{d}\delta^i_j
\hat\gamma^i_{\hphantom{i}i}= \hat\gamma^i_{\hphantom{i}j}-\frac{1}{d}\delta^i_j\theta^{\text{C}}
.
\end{equation}
The temporal connection $\hat\gamma^i_{\hphantom{i}j}$ appears also as the zero-$c$ remnant of the mixed projected relativistic Randers--Papapetrou Christoffel symbols, as in \eqref{dgammaCarlim}:
\begin{equation}
\label{dgammaCartimelim}
\frac{c}{\Omega}U_0^{\hphantom{0}\mu}\Gamma^k_{\mu\nu}h^\nu_{\hphantom{\nu}j}=\hat\gamma^{k}_{\hphantom{k}j}.
\end{equation}

The action of  $\hat \partial_t$ on scalars is simply $\partial_t$:
\begin{equation}
\label{Cartimecovdersc}
\hat \partial_t \Phi=\partial_t  \Phi,
\end{equation}
whereas on vectors or forms it
is defined as
\begin{equation}
\label{Cartimecovdervecform}
\frac{1}{\Omega}\hat \partial_tV^i=\frac{1}{\Omega} \partial_tV^i+\hat\gamma^i_{\hphantom{i}j} V^j,\quad
\frac{1}{\Omega}\hat \partial_tV_i=\frac{1}{\Omega} \partial_tV_i-\hat\gamma^j_{\hphantom{j}i} V_j.
\end{equation}
Leibniz rule generalizes the latter to any tensor and allows to demonstrate the property \eqref{cartimemet}. Indices can now be raised and lowered with the metric passing through $\hat \partial_t$.

The above Riemann--Carroll curvature tensor of a Carrollian geometry appears actually as the zero-$c$ limit of the spatial components of the ordinary Riemann curvature in the Randers--Papapetrou background.\footnote{This statement is accurate but comes without a proof. Evaluating the zero-$c$ (or infinite-$c$, as we would do in the Galilean counterpart) limit is a subtle task because in this kind of limits several components of the curvature usually diverge (see \emph{e.g.} \cite{CMPPS2}, where the r\^ole of curvature is prominent). From the perspective of the final geometry this does not produce any harm because the involved components decouple.}
In the same spirit, one may also wonder what the Carrollian limit is for the temporal components of the 
relativistic Randers--Papapetrou curvature, and this is our second and last remark. In order to answer this question, we must compute the commutator of time and space covariant derivatives acting on scalar and vector fields,
as in \eqref{carcontor} and \eqref{carriemann}. We find:
\begin{equation}
\left[\frac{1}{\Omega}\hat\partial_{t},\hat \partial_i\right]\Phi=\left(
\varphi_{i}\frac{1}{\Omega}\partial_{t}-\hat\gamma^j_{\hphantom{j}i} \hat \partial_j\right)\Phi,
\label{carriemanntime-sc}
\end{equation}
and
\begin{equation}
\left[\frac{1}{\Omega}\hat\partial_{t},\hat\nabla_i\right]V^i= \left(\hat\partial_i\theta^{\text{C}} -\hat\nabla_j\hat\gamma^{j}_{\hphantom{j}i}\right)V^i+
\left(
\theta^{\text{C}}\delta_i^j-\hat\gamma^{j}_{\hphantom{j}i}\right)\varphi_{j}
V^i+
\left(\varphi_{i}\frac{1}{\Omega}\hat\partial_{t}
-\hat\gamma^{j}_{\hphantom{j}i}\hat\nabla_j
\right)V^i
\label{carriemanntimetilde}
\end{equation}
with $\varphi_{i}$ and $\theta^{\text{C}}$ the Carrollian acceleration and expansion  \eqref{caromacc},  \eqref{carexp-tempcon}. We can define from this expression 
the components of a time-curvature Carrollian form: 
\begin{equation}
\hat r_i =\frac{1}{d}\left(\hat\nabla_j\hat\gamma^{j}_{\hphantom{j}i}
 -\hat\partial_i\theta^{\text{C}} 
\right)= \frac{1}{d}\left(\hat\nabla_j\hat\xi^{\text{C}j}_{\hphantom{\text{C}j}i}
 +\frac{1-d}{d}\hat\partial_i\theta^{\text{C}} 
\right).
\label{carriemanntime}
\end{equation}
Using $\varpi_{kl}$, $\hat r_i$ and time derivative in the framework at hand, many new curvature-like (\emph{i.e.} two-derivative) tensorial objects can be defined.
We will not elaborate any longer on these issues, which would naturally fit in a more thorough analysis of Carrollian geometry.

\subsubsection*{The Weyl--Carroll connection}

The Levi--Civita--Carroll covariant derivatives $\hat{\pmb{\nabla}}$ and $\hat\partial_t$ defined in 
\eqref{Carcovder}, \eqref{Cartimecovdersc} and \eqref{Cartimecovdervecform} for
Carrollian geometry are not covariant with respect to Weyl transformations \eqref{weyl-geometry}, 
\begin{equation}
\label{weyl-geometry-abs}
a_{ij}\to \frac{1}{\mathcal{B}^2}a_{ij},\quad b_{i}\to \frac{1}{\mathcal{B}}b_{i},\quad \Omega\to \frac{1}{\mathcal{B}}\Omega.
\end{equation} 
We can define \emph{Weyl--Carroll} covariant spatial and time derivatives using the Carrollian acceleration $\varphi_i$ defined in \eqref{caromacc} and the Carrollian expansion \eqref{carexp-tempcon},
which transform as connections (see \eqref{weyl-geometry-2}):
 \begin{equation}
 \label{weyl-geometry-2-abs}
\varphi_{i}\to \varphi_{i}-\hat\partial_i\ln \mathcal{B},\quad \theta^\text{C}\to \mathcal{B}\theta^\text{C}-\frac{d}{\Omega}\partial_t \mathcal{B}.
\end{equation} 

For a weight-$w$ scalar function  $\Phi$, \emph{i.e.} a function scaling with $\mathcal{B}^w$ under \eqref{weyl-geometry-abs},
we introduce
\begin{equation}
\label{CWs-Phi}
\hat{\mathscr{D}}_j \Phi=\hat\partial_j \Phi +w \varphi_j \Phi,
\end{equation}
such that under a Weyl transformation
\begin{equation}
\hat{\mathscr{D}}_j \Phi\to \mathcal{B}^w \hat{\mathscr{D}}_j \Phi.
\end{equation}
Similarly, for a vector with weight-$w$ components $V^l$:
\begin{equation}
\hat{\mathscr{D}}_j V^l=\hat\nabla_j V^l +(w-1) \varphi_j V^l +\varphi^l V_j -\delta^l_j V^i\varphi_i.
\end{equation}
The action on any other tensor is obtained using the Leibniz rule, as in example for rank-two tensors:
\begin{equation}
\hat{\mathscr{D}}_j t_{kl}=\hat\nabla_j t_{kl} +(w+2) \varphi_j t_{kl} +\varphi_k t_{jl}+ \varphi_l t_{kj} -
a_{jl}t_{ki}\varphi^i
-
a_{jk}t_{il}\varphi^i.
\end{equation}

The Weyl--Carroll spatial derivative does not modify the weight of the tensor it acts on.
Furthermore, it is metric as ($a_{kl}$ has weight $-2$):
\begin{equation}
\hat{\mathscr{D}}_j a_{kl}=0.
\end{equation}
It has an effective torsion because
\begin{equation}
\label{CWcontor}
\left[\hat{\mathscr{D}}_i,\hat{\mathscr{D}}_j\right]\Phi=
\frac{2}{\Omega}\varpi_{ij}\hat{\mathscr{D}}_t\Phi
+w \Omega_{ij} 
\Phi,
\end{equation}
although this expression does not contain terms of the type $\hat{\mathscr{D}}_k\Phi$. We have introduced here
\begin{equation}
\label{CWOme}
 \Omega_{ij} =\varphi_{ij} -\frac{2}{d}\varpi_{ij} \theta^\text{C},
\end{equation}
where $\varpi_{ij}$ are the components of the Carrollian vorticity defined in \eqref{caromacc}, and 
\begin{equation}
\label{carphiij}
 \varphi_{ij}=\hat\partial_i \varphi_j - \hat\partial_j \varphi_i. 
\end{equation}
Both $\Omega_{ij}$ and $ \varpi_{ij}$ are components of genuine Carrollian two-forms, and Weyl-covariant of weight $0$ and $-1$. However, $\varphi_{ij}$ are not Weyl-covariant, although they are also by construction components of a good Carrollian two-form. 

In Eq. \eqref{CWcontor}, we have used a Weyl--Carroll derivative with respect to time $\hat{\mathscr{D}}_t$. Its action on a weight-$w$ function $\Phi$ is defined as:
\begin{equation}
\label{CWtimecovdersc}
\frac{1}{\Omega}\hat{\mathscr{D}}_t \Phi=\frac{1}{\Omega}\hat\partial_t \Phi +\frac{w}{d} \theta^\text{C} \Phi=
\frac{1}{\Omega}\partial_t \Phi +\frac{w}{d} \theta^\text{C} \Phi,
\end{equation}
which is a scalar of weight $w+1$ under \eqref{weyl-geometry-abs}:
\begin{equation}
\frac{1}{\Omega}\hat{\mathscr{D}}_t \Phi \to \mathcal{B}^{w+1}\frac{1}{\Omega}\hat{\mathscr{D}}_t \Phi.
\end{equation}
Accordingly, on a weight-$w$ vector the action of the Weyl--Carroll time derivative is 
\begin{equation}
\label{CWtimecovdervecform}
\frac{1}{\Omega}\hat{\mathscr{D}}_t V^l=\frac{1}{\Omega}\hat\partial_t V^l +\frac{w-1}{d} \theta^\text{C} V^l=
\frac{1}{\Omega}\partial_t V^l +\frac{w}{d} \theta^\text{C} V^l
+\xi^{\text{C}l}_{\hphantom{\text{C}l}i} V^i .
\end{equation}
These are the components of a genuine Carrollian vector of weight $w+1$ (the tensor $\xi^{\text{C}l}_{\hphantom{\text{C}l}i}$ is Weyl-covariant of weight $1$). We have used \eqref{Cartimecovdersc}, \eqref{Cartimecovdervecform} and \eqref{carsh-tempcon} for the second equalities in \eqref{CWtimecovdersc} and \eqref{CWtimecovdervecform}. The same pattern  applies for any tensor by Leibniz rule, and in particular:
\begin{equation}
\label{CWt-met}
\hat{\mathscr{D}}_t a_{kl}=0.
\end{equation}

We will close the present appendix with the Weyl--Carroll curvature tensors, obtained by studying the commutation of Weyl--Carroll covariant derivatives acting on vectors. We find 
\begin{equation}
\label{CWcurvten}
\left[\hat{\mathscr{D}}_k,\hat{\mathscr{D}}_l\right]V^i=
\left( \hat{\mathscr{R}}^i_{\hphantom{i}jkl} - 2
\xi^{\text{C}i}_{\hphantom{\text{C}i}j}
\varpi_{kl} 
\right)
V^j+
\varpi_{kl}\frac{2}{\Omega}\hat{\mathscr{D}}_t V^i
+w \Omega_{kl}
V^i,
\end{equation}
where 
\begin{eqnarray}
\label{CWRiemann}
\hat{\mathscr{R}}^i_{\hphantom{i}jkl} &=&\hat r^i_{\hphantom{i}jkl}
-\delta^i_j\varphi_{kl}
-a_{jk} \hat{\nabla}_l \varphi^i
+a_{jl} \hat{\nabla}_k \varphi^i 
+\delta^i_k \hat{\nabla}_l \varphi_j 
-\delta^i_l \hat{\nabla}_k \varphi_j 
\nonumber
\\ 
&&+\varphi^i\left(\varphi_k a_{jl}-\varphi_l a_{jk}\right)
-\left(\delta^i_k a_{jl}-\delta^i_l a_{jk}\right)\varphi_m\varphi^m+
\left(\delta^i_k \varphi_l-\delta^i_l \varphi_k\right)\varphi_j
 \end{eqnarray}
are the components of the Riemann--Weyl--Carroll weight-$0$ tensor, from which we define
\begin{equation}
\label{CWricci-scalar}
\hat{\mathscr{R}}_{ij}=\hat{\mathscr{R}}^k_{\hphantom{k}ikj},\quad \hat{\mathscr{R}}=a^{ij}\hat{\mathscr{R}}_{ij}.
\end{equation}
Notice that the Ricci--Weyl--Carroll tensor is \emph{not} symmetric in general: $\hat{\mathscr{R}}_{ij}\neq \hat{\mathscr{R}}_{ji}$. 

Eventually, we quote
\begin{equation}
\left[\frac{1}{\Omega}\hat{\mathscr{D}}_{t},\hat{\mathscr{D}}_i\right]\Phi= w \hat{\mathscr{R}}_{i}\Phi-
\xi^{\text{C}j}_{\hphantom{\text{C}j}i}\hat{\mathscr{D}}_j^{\vphantom{j}} \Phi
\label{CWrsc}
\end{equation}
and 
\begin{equation}
\left[\frac{1}{\Omega}\hat{\mathscr{D}}_{t},\hat{\mathscr{D}}_i\right]V^i= 
(w-d)\hat{\mathscr{R}}_{i} V^i
-V^i\hat{\mathscr{D}}_j \xi^{\text{C}j}_{\hphantom{\text{C}j}i}
-\xi^{\text{C}j}_{\hphantom{\text{C}j}i}\hat{\mathscr{D}}_jV^i,
\label{CWrvec}
\end{equation}
with
\begin{equation}
\hat{\mathscr{R}}_{i}=\hat r_i
+\frac{1}{\Omega}\hat \partial_{t}\varphi_i
-\frac{1}{d} \hat\nabla_j\hat\gamma^{j}_{\hphantom{j}i}
+\xi^{\text{C}j}_{\hphantom{\text{C}j}i}\varphi_j
=\frac{1}{\Omega} \partial_{t}\varphi_i-\frac{1}{d}\left(\hat \partial_i+\varphi_i\right)\theta^\text{C}
\label{CWRvec}
\end{equation}
the components of a Weyl-covariant weight-$1$ Carrollian curvature one-form, where $\hat r_i$ is given in \eqref{carriemanntime}.

\end{document}